%% file: main.tex
\documentclass[10pt,sigconf,letterpaper,nonacm]{acmart}
\input{_util/0_package_includes}

\makeatletter
\let\hyxmp@parse@acmart\relax
\makeatother

\begin{document}

\input{_util/2_title_and_authors}
\input{sections/0_abstract}
\input{_util/3_keywords_concepts_and_ref_format}

%%%%%%%%%%%%%%%%%%%%%%%%%%%%%%%%%%%%%%%%%%%%%%%%%%%%%%

\input{sections/1_introduction}
\input{sections/3_related_work}
\input{sections/4_policy_commitments_and_timelines}
\input{sections/5_methodology}
\input{sections/6_evaluation}
\input{sections/7_discussion}
\input{sections/8_limitations}
\input{sections/9_conclusion}

\newpage
\input{sections/references}

\newpage
\input{sections/ethics}
\input{sections/appendix}

\end{document}

%% file: _util/0_package_includes.tex
\usepackage{tikz}
\usepackage{amsmath}

\usepackage{ragged2e}

\usepackage{booktabs}
\usepackage{multirow}
\usepackage[table]{xcolor}
\usepackage{makecell}

\usepackage{tcolorbox}
\usepackage{array} 
\usepackage[export]{adjustbox}

\usepackage{pifont}
\usepackage{balance}

\usepackage{hyperref}

\usepackage [english]{babel}
\usepackage [autostyle, english = american]{csquotes}
\MakeOuterQuote{"}

\newcolumntype{C}[1]{>{\centering\arraybackslash}p{#1}}
\newcolumntype{L}[1]{>{\raggedright\arraybackslash}p{#1}}
\definecolor{lightgray}{gray}{0.9} 

\newcommand{\myshape}[1]{%
  \begin{tikzpicture}[baseline=(char.base)]
    \node(char) {\phantom{X}}; % Standardize baseline to text
    \path (-0.2,0) -- (0.2,0); 
    #1
  \end{tikzpicture}%
}

\newcommand{\mycircle}[1]{%
    \myshape{\draw[line width=0.6pt, fill=#1] (0,0.015) circle (0.121);}}

\newcommand{\mysquare}[1]{%
    \myshape{\draw[line width=0.6pt, fill=#1] (-0.13,-0.1) rectangle (0.13,0.14);}}

\newcommand{\mytriangle}[1]{%
    \myshape{\draw[line width=0.6pt, fill=#1] (-0.13,-0.1) -- (0.13,-0.1) -- (0,0.13) -- cycle;}}

\usepackage[caption=false,font=footnotesize]{subfig}

\usepackage{framed}

\definecolor{lightgray}{rgb}{0.96, 0.96, 0.96}
\definecolor{darkgray}{rgb}{0.3, 0.3, 0.3}

\definecolor{lightblue}{rgb}{0.90, 0.95, 1.0}
\definecolor{boldblue}{rgb}{0.0, 0.44, 0.74}

\definecolor{lightgreen}{rgb}{0.94, 0.98, 0.94}
\definecolor{boldgreen}{rgb}{0.13, 0.55, 0.13}

\newenvironment{RQ}{

\MakeFramed{\advance\hsize-\width\FrameRestore}}
{\endMakeFramed}

\newenvironment{summary}{

\MakeFramed{\advance\hsize-\width\FrameRestore}}
{\endMakeFramed}

\newcolumntype{M}[1]{>{\centering\arraybackslash}m{#1}}

\usepackage[ruled,vlined]{algorithm2e}

%% file: _util/2_title_and_authors.tex
\title[Mind the Gap: Policy vs Reality in Post-Quantum TLS Deployment]{Mind the Gap: Policy vs Reality\\in Post-Quantum TLS Deployment}

\makeatletter
\renewcommand{\@authorfont}{\Large}
\makeatother

\author{Nimesha Wickramasinghe$^{*}$, Frank Li$^{\dagger}$, Sanjay Jha$^{*}$, Arash Shaghaghi$^{*}$}
\affiliation{%
  \institution{\textit{$^{*}$The University of New South Wales, Sydney, Australia.}\\
  \textit{$^{\dagger}$Georgia Institute of Technology,Atlanta, GA, USA}}
  \city{\endgraf}
  \country{\endgraf}
}

\renewcommand{\shortauthors}{Wickramasinghe et al.}

%% file: sections/0_abstract.tex
\begin{abstract}
Post-quantum cryptography (PQC) has evolved from a long-term planning concern into an operational priority. Following NIST’s standardization of PQC, governments and standard bodies published transition roadmaps outlining migration timelines, priority sectors, and deployment strategies. However, our survey of these policies reveals substantial divergence in technical prescriptions and urgency. It remains unclear how widely PQC has been adopted in practice and how policy differences translate into observable deployment outcomes. 

To address this gap, we present the first longitudinal measurement study of post-quantum TLS (PQ-TLS) adoption. By establishing more than 2 billion TLS handshakes, we analyze cryptographic negotiation behavior across 1 million domains from 11 globally distributed vantage points. Despite varied policy guidance, we observe configuration convergence: PQ-TLS deployment overwhelmingly centers on a single hybrid construction, and much of the apparent progress is driven by managed infrastructure providers. National timelines and sectoral priorities show limited correspondence with observed deployment patterns. Contrary to early experimental studies suggesting measurable overhead, we find that PQ-TLS introduces no meaningful latency increase in Internet settings, although it is frequently deployed alongside legacy TLS configurations. Together, these findings highlight a gap between policy expectations and early deployment reality, and provide empirical insight to inform more grounded PQ-TLS transition.
\end{abstract}

%% file: _util/3_keywords_concepts_and_ref_format.tex

\maketitle
\hypersetup{
  pdftitle={Mind the Gap: Policy vs Reality in Post-Quantum TLS Deployment},
  pdfauthor={Nimesha Wickramasinghe, Frank Li, Sanjay Jha, Arash Shaghaghi},
  pdfcreator={LaTeX with pdfTeX}
}

%% file: sections/1_introduction.tex
\section{Introduction}

The continued progress of quantum computing fundamentally challenges the security of long-standing public-key cryptography \cite{shors_algo}. In response, post-quantum cryptography (PQC) seeks to replace these vulnerable schemes with algorithms that remain secure against cryptographically relevant quantum computers (CRQCs) \cite{us_2}.

The transition to PQC has been shaped by a global standardization effort led by the National Institute of Standards and Technology (NIST). After eight years of an open, multi-round evaluation process, NIST finalized its first set of PQC algorithms in August 2024 \cite{fips_203,fips_204,fips_205}. These algorithms now form the basis for protocol development within standards bodies such as, the IETF \cite{rfc9909,ietf_tls_hybrid_design, ietf-uta-pqc-app-00} and ISO \cite{iso_sc27}, are adopted by widely used cryptographic libraries \cite{pqc_client_side_survey}, and are increasingly supported by major infrastructure providers \cite{akamai_pqc_support, aws_pqc_support}. Following these standardization efforts, governments and standards bodies worldwide have published transition roadmaps and policy guidelines to promote awareness and adoption of PQC (e.g., \cite{us_2, germany_2, aus_2}).

Across these policies, however, we observe a fragmented landscape, with substantial divergence in technical prescriptions, urgency, and deployment guidance. At the same time, the extent of PQC adoption in practice and its alignment with differing policy commitments remain largely unknown.

To address this gap, we focus on TLS, which is the most widely used cryptographic protocol on the Internet \cite{wwv_of_www}. TLS exposes cryptographic negotiation behavior directly to clients, making deployment decisions observable at Internet scale. Post-Quantum TLS (PQ-TLS), therefore, provides a practical vantage point for examining how standardized algorithms are adopted and deployed in real-world systems.

In this paper, we present the first Internet-scale longitudinal measurement study of PQ-TLS conducted after NIST standardization. Our study spans three measurement rounds, conducted in July 2025, November 2025, and March 2026, with approximately four months between each round. We employ a custom TLS~1.3 measurement client and conduct controlled handshake probing from 11 globally distributed vantage points. In total, we analyze cryptographic negotiation of more than 2 billion TLS handshakes spanning one million domains, enabling direct observation of how PQ-TLS adoption evolves over time.

Our measurements reveal that despite diverse policy guidance and a broad standardized design space, operational deployment rapidly converges on a single hybrid configuration, \texttt{X25519MLKEM768}. Observed adoption is overwhelmingly driven by managed infrastructure providers (e.g., CDNs), with nearly 70\% of PQ-TLS deployment attributable to two platforms, Cloudflare and Fastly. National timelines and sectoral priorities show limited correspondence with observed deployment patterns. Moreover, contrary to early experimental expectations of performance overhead, we find that PQ-TLS introduces no measurable latency or compatibility penalties at Internet scale, although PQ mechanisms are frequently deployed alongside legacy TLS configurations.

Overall, these findings highlight a disconnect between policy expectations and early deployment realities, while offering empirical evidence to inform more grounded PQC transition strategies.

%% file: sections/3_related_work.tex
\section{Background and Related Work}
\label{sec:literature_review}

This section summarizes the role of PQ mechanisms in TLS and reviews prior work on TLS measurement, early PQC deployment, and performance evaluation.

\subsection{Background}
PQ-TLS extends the TLS~1.3 handshake to incorporate PQ primitives for key establishment and authentication, replacing classical public-key mechanisms with PQ alternatives \cite{ietf_tls_hybrid_design}. A concise overview of TLS cryptographic components, the quantum threat model, and the NIST PQC standardization effort is provided in Appendix~\ref{appendix:background} to contextualize the deployment measurements presented in this work.

\subsection{Related Work}
Internet-scale measurement has long provided the empirical basis for understanding how TLS evolves on the public Internet. Early work \cite{ssl_landscape} mapped the SSL/TLS ecosystem using active and passive measurements, revealing concentration and misconfiguration in the Web PKI. \cite{https_certificate_ecosytem} subsequently provided a longitudinal view of the HTTPS certificate ecosystem, while \cite{largest_security_study} examined the evolution of HTTPS deployment at scale. Complementary work \cite{weak_keys_remain} demonstrated that weak cryptographic keys remained widespread in network devices years after disclosure, illustrating the persistence of long-tail security failures. More recent studies have tracked TLS deployment, including longitudinal adoption trends~\cite{coming_of_age}, the rollout and centralization dynamics of TLS~1.3~\cite{tracking_the_deployment_of_tls, www-21-tls-1.3-audit}, and regional variation in negotiated TLS behavior~\cite{geo_differences_in_tls}. Our work builds on these measurement methodologies and extends them to study the emerging PQC transition.

A growing body of work has examined PQC readiness from the perspectives of client support, cryptographic library integration, and provider telemetry~\cite{pqc_client_side_survey, cloudflareDataExplorer}. Experimental deployments such as CECPQ1 and CECPQ2 demonstrated the feasibility of hybrid PQ key exchange prior to standardization~\cite{CECPQ1, CECPQ2}. Controlled studies have evaluated PQ-TLS performance in simulated settings and testbeds~\cite{pqc_performance_2, pqc_performance, performance_controlled_1, performance_controlled_2, performance_controlled_3}, while vendor-specific reports offer insights into deployment within individual ecosystems~\cite{cloudflare_experiment_1}. 

To the best of our knowledge, this study presents the first longitudinal, vendor-agnostic study of publicly deployed PQ-TLS after NIST standardization, and the first to explicitly connect these observations to national PQC transition policies. Our contribution is therefore not only to assess whether PQ-TLS is technically viable, but also to explain \textit{how}, \textit{where}, and \textit{by whom} it is being deployed on the public Internet. 

%% file: sections/4_policy_commitments_and_timelines.tex
\section{Research Questions}
\label{sec:policy_commitments}

\input{tables/1_pqc_policies}

To ground our measurement study in comparable policy contexts, we survey publicly available national and regional PQC transition documents. 

We construct our policy set by selecting countries with demonstrated engagement in PQC, including participants in NIST’s standardization process  \cite{nist_first_round,ship_has_sailed}, and major economic regions \cite{world_bank}. For each country or region, we collect published policies through systematic web-based searches and retain documents from authoritative sources, including government publications and national standards bodies. Full selection methodology and inclusion criteria are provided in Appendix~\ref{appendix:shortlisted_countries}.

Table~\ref{tab:1_pqc_policies} summarizes the selected policies and the dimensions most relevant to PQ-TLS deployment, including recommended algorithms, guidance on hybrid constructions, sectoral scope, and transition timelines. While not exhaustive, our survey of eight countries captures a cross-section of PQC transition approaches whose differing requirements may place varying demands on protocol implementers, infrastructure providers, and service operators. 

To assess how these policy expectations translate into observable deployment behavior, and whether they converge or diverge at the protocol level, we formulate a set of targeted research questions aligned with each policy dimension.

\subsection{Algorithm Recommendations}
Across the surveyed policies, there is clear convergence on ML-KEM as the primary PQ replacement for classical key-exchange in TLS. However, this consensus masks meaningful divergence in deployment guidance. France and Germany additionally reference alternative KEMs such as Frodo-KEM or Classic McEliece, while other jurisdictions focus exclusively on ML-KEM. Parameter recommendations also vary: Australia, India, and the United States of America (USA) explicitly prefer higher security levels (e.g., ML-KEM-1024), Canada permits a broader range of parameters, and France and Germany leave parameter selection largely unspecified.

Guidance on PQ signature algorithms is less prescriptive. Although ML-DSA and SLH-DSA are commonly referenced, some policies additionally mention XMSS or LMS.

Hybrid constructions, which combine a classical cryptographic algorithm (e.g., X25519) with a PQ algorithm (e.g., ML-KEM), are widely discussed but positioned differently across jurisdictions. Australia and the USA permit hybrid deployment without explicitly endorsing it, the United Kingdom frames hybrid use as transitional, and other policies recommend hybrid constructions without constraints.

Taken together, the diversity in algorithms, security levels, and hybrid constructions motivates an empirical examination of how PQ-TLS is deployed in practice.

\begin{RQ}
\noindent\textbf{RQ1 – Configuration Landscape:}
\textit{Which standardized PQ-TLS configurations are deployed in practice, and how does the observed configuration landscape compare to the diversity articulated in national transition policies?}
\end{RQ}

\subsection{Infrastructure Concentration}
National PQC transition policies are framed with limited explicit consideration of the infrastructure platforms. In practice, however, a substantial fraction of Internet traffic is served by large-scale CDNs and managed hosting providers \cite{www-21-tls-1.3-audit, tracking_the_deployment_of_tls, uncovering_hosting_ips, formalizing_depenedence_of_web_infrastructure}.

Deployment decisions made at the infrastructure layer can therefore exert disproportionate influence over observed adoption patterns, security posture, and performance characteristics. Distinguishing between infrastructure-driven rollout and independent operator migration is essential for understanding whether observed PQC adoption reflects systemic readiness or infrastructure-level defaults.

\begin{RQ}
\noindent\textbf{RQ2 - Deployment Agency:}
\textit{How much of observed PQ-TLS adoption is driven by infrastructure-managed deployment versus potentially owner-managed migration?}
\end{RQ}

\subsection{Policy Scope and Timelines}
The surveyed policies differ in both scope and transition timelines, defining which sectors should prioritize PQC adoption and when migration should be completed. Some jurisdictions, including Australia and the USA, explicitly mandate migration for government systems, national security, and critical infrastructure, while others provide broader advisory guidance spanning public and private sectors. These differences reflect varying national strategic priorities, particularly for systems handling sensitive or long-lived data.

Timelines likewise diverge. As summarized in Table~\ref{tab:1_pqc_policies}, several jurisdictions, including Australia, Canada, and the European Union, target completion of PQC migration by 2030–2031, whereas others extend transition to 2035. 

Such variation creates measurable expectations: earlier deadlines and narrower sectoral mandates could plausibly correspond to higher present-day readiness.

Internet-facing services span diverse sectors, including finance, healthcare, telecommunications, government services, and consumer platforms, each with differing regulatory exposure and constraints. Examining PQ-TLS deployment across countries and sectors therefore enables an empirical assessment of whether policy-defined scope and timelines manifest in observable deployment differences.

\begin{RQ}
\noindent\textbf{RQ3 – Policy Alignment:}
\textit{To what extent do national transition timelines and sectoral prioritizations correspond to observable differences in PQ-TLS deployment?}
\end{RQ}

\subsection{Operational Characteristics}
Early ecosystem-specific measurements have suggested that PQ-TLS introduces measurable performance overhead during deployment. For example, Google reported that the larger key share size of Kyber resulted in a 4\% median increase in TLS handshake latency in Chrome on desktop \cite{google_pqc_latency_2}. Such findings highlight operational costs associated with larger key sizes and handshake messages in PQ constructions \cite{pqc_barriers_1}.

National transition policies, however, provide limited discussion of deployment constraints such as performance overhead, middlebox compatibility, or handshake reliability at Internet scale. Implicit in many policy roadmaps is the assumption that standardized PQ algorithms can be integrated into operational systems without prohibitive impact.

For Internet-facing security protocols, operational viability encompasses not only cryptographic computation cost but also handshake latency, message size expansion, and negotiation reliability across heterogeneous infrastructures. Empirical measurement of these characteristics is therefore essential to ensure that policy recommendations are compatible with real-world deployment constraints.

\begin{RQ}
\noindent\textbf{RQ4 – Operational Viability:}
\textit{Does PQ-TLS introduce measurable performance, reliability, or compatibility constraints in operational Internet deployments?}
\end{RQ}
\vspace{-5pt}

\subsection{Security Co-evolution}
While the surveyed policies provide guidance on \emph{which} cryptographic algorithms to adopt and \emph{when} to transition, they focus primarily on primitive-level replacement. \emph{How} these algorithms should be integrated into Internet-facing protocols such as TLS is typically left implicit or deferred to existing security baselines and operational practices, which may evolve independently of PQ transition planning (e.g., \cite{uk_3, india_3}).

However, prior Internet-scale measurement studies have consistently shown that protocol-level security depends not only on cryptographic primitives but also on configuration hygiene, including supported protocol versions, cipher suites, certificate management practices, and exposure to legacy vulnerabilities \cite{stale_tls_certificates, cipher_consequences, imperfect_fs}. 

As PQ-TLS deployment progresses, an open question is whether PQ adoption occurs alongside broader security modernization, or whether PQ mechanisms are layered onto existing configurations without corresponding improvements in overall protocol security. Empirical evaluation of this relationship is necessary to understand whether observable PQ adoption reflects holistic security advancement or isolated cryptographic substitution.

\begin{RQ}
\noindent\textbf{RQ5 – Security Co-evolution:}
\textit{Does PQ-TLS adoption coincide with broader improvements in protocol-level security posture?}
\end{RQ}

%% file: tables/1_pqc_policies.tex
% \usepackage{booktabs}
% \usepackage{multirow}
\begin{table*}[!t]
\renewcommand{\arraystretch}{0.8}

    \caption{Comparative summary of national and regional PQC transition policies relevant to TLS}
    \label{tab:1_pqc_policies}
    
    \centering

    \resizebox{\textwidth}{!}{%
        \rowcolors{2}{}{lightgray}    
        
        \begin{tabular}{lcccccccc}
        
        \toprule
          
          \multicolumn{1}{c}{\multirow{2}{*}{\textbf{Country / Region}}} &
          
          \multicolumn{2}{c}{\textbf{Recommended PQC Algorithms}} &
          \multicolumn{1}{c}{\multirow{2}{*}{\textbf{Hybrid Stance}}} &

          \multicolumn{1}{c}{\multirow{2}{*}{\begin{tabular}[c]{@{}c@{}}\textbf{Requirement}\\\textbf{\& Sector}\end{tabular}}} &
          
          \multicolumn{2}{c}{\textbf{PQC Transition Timeline}} &
          \multicolumn{2}{c}{\textbf{Classical Crypto. Timeline}} \\
          % \multicolumn{1}{c}{\multirow{2}{*}{\begin{tabular}[c]{@{}c@{}}\textbf{Protocol}\\\textbf{Hygiene}\end{tabular}}} \\ 
          
          % \cmidrule(lr){2-3} \cmidrule(lr){6-9} \cmidrule(lr){10-11} 
          \cmidrule(lr){2-3} \cmidrule(lr){6-7} \cmidrule(lr){8-9} 
            
          \multicolumn{1}{c}{} &
          
          \multicolumn{1}{c}{\textbf{KEM}} &
          \multicolumn{1}{c}{\textbf{SA}} &
          \multicolumn{1}{c}{} &

          \multicolumn{1}{c}{} &
          
          % \multicolumn{1}{c}{\textbf{Plan by}} &
          % \multicolumn{1}{c}{\textbf{Start by}} &
          \multicolumn{1}{c}{\textbf{Complete by}} &
          \multicolumn{1}{c}{\textbf{Default by}} &
          \multicolumn{1}{c}{\textbf{Deprecated by}} &
          \multicolumn{1}{c}{\textbf{Disallow by}} \\ 
          
          \midrule
        
        Australia \cite{aus_1,aus_2}  &
          \multicolumn{1}{c}{\begin{tabular}[c]{@{}c@{}}ML-KEM - 768/1024\\ Prefer 1024\end{tabular}} &
          \multicolumn{1}{c}{\begin{tabular}[c]{@{}c@{}}ML-DSA - 65/87\\ 65 only till 2030\end{tabular}} &
          \multicolumn{1}{c}{\begin{tabular}[c]{@{}c@{}}Allowed\\Not recommended\end{tabular}} &
          \mycircle{black} \mysquare{black} \mytriangle{none} &
          % 2026 &
          % 2028 &
          2030 &
          2030 &
          2030 &
          2030 \\
          % \checkmark \\
    
          \cmidrule(lr){1-9}

        \rowcolor{white}
        Canada \cite{canada_1,canada_2}  &
          \multicolumn{1}{c}{\begin{tabular}[c]{@{}c@{}}ML-KEM - 512/768/1024\end{tabular}} &
          \multicolumn{1}{c}{\begin{tabular}[c]{@{}c@{}}ML-DSA - 44/65/87\\SLH-DSA\end{tabular}} &
          \multicolumn{1}{c}{\begin{tabular}[c]{@{}c@{}}NS\end{tabular}} &
          \mycircle{none} \mysquare{none} \mytriangle{none} &
           % 2026 &
           % 2026 &
           2031 &
           2035 &
           2031 &
           2031 \\
           % NS \\
    
          \cmidrule(lr){1-9}

        Europe \cite{eu_1}  &
          \multicolumn{1}{c}{\begin{tabular}[c]{@{}c@{}}Promote international\\cooperation\end{tabular}} &
          \multicolumn{1}{c}{\begin{tabular}[c]{@{}c@{}}Promote international\\cooperation\end{tabular}} &
          Recommended &
          \multicolumn{1}{c}{\begin{tabular}[c]{@{}c@{}}Advisory for\\member states\end{tabular}} &
          % 2026 &
          % NS &
          2030 &
          2035 &
          2035 &
          NS \\
           % \\
    
          \cmidrule(lr){1-9}
          
       \rowcolor{white}
        France \cite{france_1} &
          \multicolumn{1}{c}{\begin{tabular}[c]{@{}c@{}}Kyber (ML-KEM)\\ Frodo-KEM\end{tabular}} &
          \multicolumn{1}{c}{\begin{tabular}[c]{@{}c@{}}Dilithium (ML-DSA)\\ FN-DSA, XMSS, LMS\end{tabular}} &
          Recommended &
          \mycircle{none} \mysquare{none} \mytriangle{none} &
          % NS &
          % NS &
          2030 &
          NS &
          NS &
          NS \\
           % \\
    
          \cmidrule(lr){1-9}

        Germany \cite{germany_1, germany_2} &
          \multicolumn{1}{c}{\begin{tabular}[c]{@{}c@{}}ML-KEM, Frodo-KEM\\Classic McEliece\end{tabular}} &
          \multicolumn{1}{c}{\begin{tabular}[c]{@{}c@{}}ML-DSA - 65/87\\SLH-DSA, XMSS, LMS\end{tabular}} &
          Recommended &
          \mycircle{none} \mysquare{none} \mytriangle{none} &
          % NS &
          % NS &
          2030 &
          NS &
          NS &
          NS \\
           % \\
    
          \cmidrule(lr){1-9}

        \rowcolor{white}
        India \cite{india_1,india_2} &
          \multicolumn{1}{c}{\begin{tabular}[c]{@{}c@{}}ML-KEM - 768/1024\\Prefer 1024\end{tabular}} &
          \multicolumn{1}{c}{\begin{tabular}[c]{@{}c@{}}ML-DSA - 65/87\\ SLH-DSA\end{tabular}} &
          Recommended &
          \mycircle{none} \mysquare{none} \mytriangle{none} &
          % NS &
          % \multicolumn{1}{c}{\begin{tabular}[c]{@{}c@{}}Immediate\\(2025)\end{tabular}}  &
          NS &
          2035 &
          NS &
          NS \\
           % \\
    
          \cmidrule(lr){1-9}
           
        United Kingdom \cite{uk_1, uk_2} &
          \multicolumn{1}{c}{\begin{tabular}[c]{@{}c@{}}ML-KEM - 768\end{tabular}} &
          \multicolumn{1}{c}{\begin{tabular}[c]{@{}c@{}}ML-DSA - 65\\SLH-DSA, XMSS, LMS\end{tabular}} &
          Transitional &
          \mycircle{none} \mysquare{none} \mytriangle{none} &
          % 2028 &
          % 2031 &
          2035 &
          2035 &
          NS &
          NS \\
          % NS \\
    
          \cmidrule(lr){1-9}

        \rowcolor{white}
        USA \cite{us_1_nsm_10, us_2} &
          \multicolumn{1}{c}{\begin{tabular}[c]{@{}c@{}}ML-KEM - 1024\end{tabular}} &
          \multicolumn{1}{c}{\begin{tabular}[c]{@{}c@{}}ML-DSA - 87\end{tabular}} &
          \multicolumn{1}{c}{\begin{tabular}[c]{@{}c@{}}Allowed\\Not recommended\end{tabular}} &
          \mycircle{black} \mysquare{black} \mytriangle{none} &
          % NS &
          % 2026 &
          2035 &
          NS &
          2030 &
          2035 \\
          % Explicit \\

          \bottomrule

          \addlinespace[10pt]
            \multicolumn{9}{l}{ 

                \begin{tabular}[l]{@{}l@{}}

                    \rowcolor{white}
                    \mycircle{black}/\mycircle{none} \textbf{(circle)}: Government / National Security; \hspace{0.4cm} 
                    \mysquare{black}/\mysquare{none} \textbf{(square)}: Critical / high-risk / long-lived systems; \hspace{0.4cm} 
                    \mytriangle{black}/\mytriangle{none} \textbf{(triangle)}: Industry / organisations; \\[3pt]

                    \textbf{Filled} (\mycircle{black} \mysquare{black} \mytriangle{black}): Mandatory; \hspace{0.4cm} 
                    \textbf{Hollow} (\mycircle{none} \mysquare{none} \mytriangle{none}): Advisory; 
                    \hspace{0.4cm} 
                    \textbf{NS}: Not Specified\\[3pt]
                    
                \end{tabular}
                
            }
          
     \end{tabular}
}
\end{table*}

%% file: sections/5_methodology.tex
\section{Methodology}
\label{sec:methodology}
This section outlines the domain selection process (\S~\ref{subsec:domain_selection}), measurement infrastructure (\S~\ref{subsec:measurement_infra}), TLS probing techniques (\S~\ref{subsec:probing}), TLS management attribution (\S~\ref{subsec:tls_management_attribution}), and longitudinal measurement rounds (\S~\ref{subsec:longitudinal_tracking}) used to analyze PQ-TLS adoption across the Internet.

\subsection{Target Selection and Domain Coverage}
\label{subsec:domain_selection}
Our measurements are based on DomCop’s Top~10M Domains list \footnote{\url{https://www.domcop.com/top-10-million-domains}}, which is primarily derived from Open PageRank data. The open-source initiative, Open PageRank, estimates the relative importance of domains using web graph data from Common Crawl \cite{commoncrawlCommonCrawl}.

For configuration landscape (RQ1), deployment agency (RQ2), operational viability (RQ4), and security co-evolution (RQ5) analyses, we focus on the Top~1M domains. This subset captures a broad cross-section of widely deployed Internet services while maintaining feasibility for repeated longitudinal measurement. 

To support country level policy alignment analysis (RQ3), we additionally draw from the broader Top~10M corpus to construct per-country samples, including Top~10K nationally relevant domains and Top~1K government-operated domains.

We select DomCop for two methodological reasons. First, many country-specific government and public-sector domains do not appear in global Top~1M lists (e.g., Tranco \cite{tranco}), despite being operationally significant within their jurisdictions. Access to a larger corpus therefore enables the construction of more representative national samples. Although the Google Chrome UX Report (CrUX) \cite{crux} provides broader coverage, it reports domains in coarse rank-order magnitude buckets (i.e., Top 1K, 10K, 100K, 1M, and >1M) rather than exact rankings, limiting the ability to systematically select top domains \cite{crux_issues}. Second, using a single, consistent ranking source across both global and country level measurements ensures internal consistency across our analyses.

To assess the robustness of our findings, we replicate PQ-TLS adoption measurements using widely used top 1M domain rankings, including Tranco and CrUX. The adoption rates are comparable and are reported in Appendix~\ref{appendix:tranco_crux}. 

\subsection{Regional Scanning Infrastructure}
\label{subsec:measurement_infra}
Content delivery networks, geo-replicated cloud services, and regulatory environments can result in region-dependent TLS behavior \cite{geo_differences_in_tls}. To capture this diversity, we conducted measurements from 11 geographically distributed vantage points: North America West (USA, California), North America Central (Canada, Montreal), North America East (USA, Virginia), South America (Brazil, São Paulo), Africa (South Africa, Cape Town), United Kingdom (England, London), Europe (Germany, Berlin), Middle East (United Arab Emirates, Dubai), South Asia (India, Mumbai), East Asia (Japan, Tokyo), and Oceania (Australia, Sydney). 

All vantage points were hosted on cloud infrastructure and operated independently, allowing us to observe region-specific negotiation outcomes. 

To ensure consistency and mitigate measurement bias, all scans were conducted from dedicated cloud instances provisioned with identical CPU, memory, and network configurations. Scanning tools were deployed from identical precompiled binaries to avoid client-side fingerprinting discrepancies and ensure uniform TLS behavior across vantage points. Measurements were executed within shared time windows to reduce temporal variance and limit the impact of configuration changes during the measurement period.

\subsubsection{DNS Resolution and Regional Targeting}
Many domains are served by CDNs or geo-replicated infrastructures that return different IP addresses depending on client location. To account for this behavior, we perform DNS resolution locally at each vantage point prior to scanning. For each domain, we resolve all available A and AAAA records and apply a latency-based IP selection strategy following prior work \cite{www-21-tls-1.3-audit}. Specifically, we measure the round-trip time to each candidate IP using TCP connection attempts on port 443 and select the address with the lowest observed latency among responsive endpoints. We then perform TLS scans against that endpoint, using the domain name in the SNI extension. This approach ensures that our measurements reflect the endpoint most representative of the user experience in each region.

\subsection{Scanning Tools and Measurements}
\label{subsec:probing}

We develop a custom TLS scanning client, which provides native support for NIST-standardized PQC algorithms. The client is configured to resemble a security-conscious modern TLS~1.3 implementation, explicitly excluding support for legacy protocol versions.

For each target domain, we first establish two baseline TLS~1.3 handshakes to identify the server’s default cryptographic preferences. The first is an \emph{unconstrained} handshake in which the client advertises all supported classical, hybrid, and PQ algorithms. This reveals the server’s \textbf{default} key exchange ($\textit{default\_KE}$) and default signature ($\textit{default\_SA}$) algorithms under normal client conditions. The second is a \emph{classical-only} handshake in which the client advertises only classical algorithms, yielding the server’s default classical key exchange (\textit{default\_classical\_KE}) and default classical signature algorithm ($\textit{default\_classical\_SA}$) when PQ options are unavailable.

Using these preferences as anchors, we probe the server’s \textbf{supported} cryptographic capabilities without enumerating the full Cartesian product of key exchange and signature algorithms. To this end, we fix the key exchange group to ($\textit{default\_KE}$) and iterate over supported PQ signature algorithms. Conversely, we fix the signature algorithm to ($\textit{default\_SA}$), we iterate over all key exchange groups supported by the client to determine the server’s supported key establishment mechanisms.

We measure the operational impact of PQC components by applying a differential measurement strategy that isolates the performance contribution of key exchange and signature algorithms independently. To assess the performance impact of PQ \emph{signature algorithms}, we restrict our analysis to domains that advertise support for at least one PQ signature scheme. For each such domain, we fix the client supported key exchange group to $\textit{default\_classical\_KE}$, and iterate over the server's supported PQ signature algorithms. For every evaluated configuration, we immediately perform a corresponding baseline handshake using $\textit{default\_classical\_KE}$ and $\textit{default\_classical\_SA}$, allowing any observed differences in latency or message size to be attributed solely to the signature algorithm. Similarly, to assess the performance impact of PQ key exchange mechanisms, we restrict measurements to domains that support hybrid or pure PQ key exchange. For these domains, we fix the signature algorithm to $\textit{default\_classical\_SA}$ and iterate over the server's supported hybrid and pure key exchange groups. For hybrid key exchange groups, we further control for the classical component by selecting the corresponding classical key exchange as the baseline (e.g., comparing \texttt{X25519MLKEM768} against \texttt{X25519}), provided the classical counterpart is supported by the server. This design isolates the performance impact attributable to the PQ component of each construction. For pure PQ key exchanges, we compare against an immediate baseline handshake using $\textit{default\_classical\_KE}$ and $\textit{default\_classical\_SA}$. 

For each PQ–enabled configuration, we record the TLS handshake time as well as the total number of bytes transmitted and received. Let $T_{\text{pqc}}(D, c)$ denote the handshake time for domain $D$ under a PQ configuration $c$, and $T_{\text{classical}}(D, c')$ the time for the corresponding classical baseline $c'$. We define the latency delta as $\Delta_T(D, c) = T_{\text{pqc}}(D, c) - T_{\text{classical}}(D, c')$. Analogously, we compute byte-level deltas for transmitted and received handshake data as $\Delta_{B^{\text{tx}}}(D, c)$ and $\Delta_{B^{\text{rx}}}(D, c)$.

For each evaluated handshake configuration, the client establishes five matched PQ and baseline TLS handshake pairs using identical parameters, spacing successive pairs by four hours. Reporting results across multiple handshakes reduces sensitivity to transient network jitter and short-lived congestion effects. The enforced four-hour spacing further minimizes the influence of localized traffic spikes, routing instability, or load-balancing artifacts, ensuring that observed differences primarily reflect protocol behavior rather than measurement noise. Accordingly, we restrict measurements to at most one scan pair (PQ and baseline-classical) per domain every four hours for any given configuration. 

\subsection{TLS Management Attribution}
\label{subsec:tls_management_attribution}

To examine whether observed PQ-TLS adoption reflects independent operator migration or infrastructure-driven deployment, we attribute each domain to the entity responsible for terminating its TLS connections and subsequently infer who is potentially responsible for managing its TLS configuration.

We first identify the infrastructure provider behind each domain using a layered resolution strategy. We begin by matching the resolved IP address against known CDN and cloud provider ranges (e.g., Cloudflare\footnote{\url{https://www.cloudflare.com/ips-v4}}, Fastly\footnote{\url{https://api.fastly.com/public-ip-list}}). If no match is found, we analyze DNS CNAME chains, reverse DNS (PTR) lookups, HTTP(S) response metadata, and WHOIS records for provider-specific patterns. When these signals are inconclusive, we use autonomous system (AS) ownership information as a fallback.

\begin{sloppypar}
Given the inferred infrastructure provider, we then estimate whether TLS configuration is managed by the provider or by the domain operator following \cite{www-21-tls-1.3-audit}. Domains associated with known managed infrastructure platforms (e.g., CDNs), associated with platform-as-a-service offerings (e.g., github.io, blogspot.com), or exhibiting characteristics of shared hosting such as high IP fanout and known infrastructure-linked metadata, are labeled as \emph{potentially infrastructure provider managed}. In contrast, domains with characteristics of dedicated hosting, including low fanout, domain-aligned certificates are labeled as \emph{potentially owner managed}. Domains that cannot be reliably classified are labeled as \emph{unknown}.

This framework provides probabilistic attribution, enabling us to distinguish between deployments \emph{potentially} driven by infrastructure platforms and those managed directly by service operators.
\end{sloppypar}

\subsection{Longitudinal Tracking}
\label{subsec:longitudinal_tracking}
To enable longitudinal analysis of PQ-TLS deployment, we anchor all measurements to DomCop Top~10M domains obtained in July 2025. This identical domain set is reused for scans conducted in July 2025, November 2025, and March 2026, allowing direct observation of changes in default cryptographic negotiation behavior over time without confounding effects from ranking churn or domain turnover.

A domain is treated as unsuccessful if it fails to complete a TLS handshake after three retry attempts.

%% file: sections/6_evaluation.tex
\section{Evaluation}
\label{sec:policy_to_practice}

\input{tables/0_pqc_adoption}

In this section, we provide an empirical evaluation of how current Internet-facing TLS deployments align with the policy commitments and transition timelines surveyed in \S~\ref{sec:policy_commitments}.

\vspace{0.25cm}\noindent\textbf{\textit{TLS Handshake Outcomes}} \hfill\\
We define a \emph{successful scan} as a domain that completes an unconstrained TLS~1.3 handshake from every measurement vantage point in all three measurement rounds (July~2025, November~2025, and March~2026). Under this definition, 684{,}494 domains (68.44\%) consistently negotiated TLS~1.3 handshakes across all regions and time points. These domains constitute the \textbf{stable panel} used for longitudinal deployment analysis.

The remaining 315{,}506 domains (31.55\%) failed to complete an unconstrained TLS 1.3 handshake from at least one vantage point in at least one measurement round, despite three retry attempts. The majority of these failures consistently stem from TLS version mismatches, reflecting servers that do not successfully negotiate TLS 1.3, as well as DNS resolution, and transport-layer issues. We further examine these failure modes and their underlying causes in \S~\ref{subsubsec:failure_modes}.

\subsection{Configuration Landscape}
\label{subsec:rq1_adoption}

Across the stable panel, \textbf{we do not observe any successful negotiation of PQ signature algorithms}. This absence is consistent across all vantage points and measurement rounds, indicating that PQ signature deployment remains negligible in publicly reachable TLS services during the study period. This finding aligns with the urgent transition focus on PQ key establishment relative to authentication (Appendix~\ref{subapp:pqc_urgency}). Consequently, the remainder of our analysis focuses on PQ key establishment, which represent the only observable form of PQC deployment in operational TLS at present.

\input{figures/1_infra_centralisation/1_1_config_attribution_figure}
\input{figures/1_infra_centralisation/2_infra_providers}

\begin{sloppypar}
Table~\ref{tab:0_pqc_adoption} characterizes the PQ key establishment deployment among the stable panel. The adoption increases markedly over the eight-month study period. In July~2025, 31.26\% of domains negotiate a hybrid PQ key exchange group by default. This rises to 47.37\% ($\uparrow 16.11$) in November~2025 and further to 49.22\% ($\uparrow 1.85$) in March~2026. By the end of the measurement period, \textbf{nearly half of the stable domain panel defaults to hybrid PQ key exchange}, while the remainder continues to negotiate classical groups. 

Despite this rapid adoption, the configuration landscape exhibits minimal diversity. Across all measurement rounds, \textbf{every domain that negotiates PQ key exchange by default selects the same hybrid construction, \texttt{X25519MLKEM768}}. Alternative hybrid constructions appear only as secondary capabilities. By March~2026, 5.53\% of domains advertise support for \texttt{secp256r1MLKEM768} and 1.95\% for \texttt{secp384r1MLKEM1024}, yet neither configuration is ever observed as the negotiated default.
\end{sloppypar}
Examining the number of PQ configurations supported per domain further highlights this concentration. By March~2026, 336{,}919 domains advertise support for at least one PQ key exchange mechanism. Among these, 88.76\% support exactly one PQ configuration, exclusively \texttt{X25519MLKEM768}. An additional 7.28\% support two configurations and 3.96\% support three, but in all such cases the supported options remain hybrid constructions. \textbf{Pure \texttt{ML-KEM-*} key exchange groups are not deployed in isolation}, and only three domains advertise a broader set of PQ options that includes both hybrid and pure constructions.

Beyond aggregate adoption, the longitudinal measurements also reveal a high degree of deployment stability. In early measurements we observe a small number of domains that advertise support for PQ key exchange but continue to negotiate classical groups by default. Specifically, in July~2025, 214{,}093 domains advertise PQ capability while 214{,}020 negotiate PQ by default, leaving 73 domains that retain PQ support without prioritizing it. However, this gap disappears in subsequent measurement rounds. By March~2026, only two domains exhibit this behavior. Moreover, we do not observe any instances where domains that previously negotiated PQ key exchange revert to classical-only configurations. These observations suggest that \textbf{PQ-TLS rollout proceeds in a largely monotonic manner: once enabled and promoted to the default configuration, deployments tend to remain stable}.

\begin{summary}
\noindent\textbf{Key Takeaways for RQ1:}
\textit{Despite diverse policy recommendations, operational PQ-TLS has effectively standardized on a single hybrid construction. Observed adoption reflects a narrow set of interoperable default configurations, rather than the broader range of algorithmic options described in PQC transition guidelines.}
\end{summary}

\subsection{Deployment Agency}
\label{subsec:rq2_infrastructure}
To assess whether observed PQ-TLS adoption reflects independent operator migration or infrastructure-driven deployment, we apply the TLS management attribution methodology described in \S~\ref{subsec:tls_management_attribution}. Based on this analysis, we classify each domain as \emph{potentially infrastructure provider managed}, \emph{potentially owner managed}, or \emph{unknown}, reflecting the inferred responsibility for TLS configuration. 

Among the 684{,}494 domains in the stable panel, 416{,}669 (60.87\%) are attributed to potentially infrastructure provider managed deployments, 205{,}960 (30.09\%) are potentially owner managed, and 61{,}865 (9.04\%) remain unattributed. 

By March~2026, \textbf{93.92\% of observed PQ-TLS deployments originate from infrastructure provider managed configurations}. In contrast, \textbf{only 4.58\% of observed PQ-TLS deployments were owner managed}, with the majority continuing to negotiate classical TLS. 

\begin{sloppypar}
Figure~\ref{fig:1_1_config_attribution} further illustrates this imbalance within attribution categories. More than 75\% of infrastructure provider managed domains support PQ-TLS, compared to less than 10\% of owner managed domains. Taken together, these disparities indicate that the \textbf{majority of observable PQ-TLS adoption is driven by platform-level configuration changes rather than deliberate migration by individual service operators}.
\end{sloppypar}

To further examine the role of infrastructure providers, Figure~\ref{fig:2_infrastructure_providers}-(a) presents the distribution of PQ-TLS–default domains within the providers responsible for terminating their TLS connections. PQ-TLS deployments are highly concentrated among a small number of platforms. In particular, \textbf{Cloudflare and Fastly together account for nearly 70\% of all observed PQ-TLS adoption}.

Figure~\ref{fig:2_infrastructure_providers}-(b) shows how PQ-TLS adoption evolves within these infrastructure providers over the course of the measurement period. Early adoption is primarily driven by Cloudflare, Fastly and Squarespace, with a substantial portion of their hosted domains already negotiating PQ-TLS in July~2025. Subsequent measurement rounds show incremental adoption across additional platforms, including Amazon through CloudFront, Squarespace, Automattic, and others. Given the active role of these providers in experimenting with and deploying PQ-TLS, their prominence among PQ-enabled domains is expected. However, \textbf{even within infrastructure providers, a considerable fraction of served domains continue to negotiate classical-TLS} as of March~2026, indicating that provider-level rollouts currently remain in progress.

\begin{summary}
\noindent\textbf{Key Takeaways for RQ2:}
Observed PQ-TLS adoption is largely driven by infrastructure-provider deployment rather than deliberate migration by individual domain operators. A small number of large-scale platforms account for the majority of PQ-enabled domains, highlighting the central role of infrastructure providers in translating policy expectations into observable deployment.
\end{summary}

\subsection{Policy Alignment}
\label{subsec:rq3_policy_alignment}

\input{figures/2_sectors_and_countries/2_policy_alignment_figure}

We assess whether observed PQ-TLS deployment aligns with the priorities and timelines outlined in PQC transition strategies by analyzing trends across policy-relevant sectors (\S~\ref{subsubsec:sectoral_adoption}) and countries (\S~\ref{subsubsec:national_adoption}).

\subsubsection{Sectoral Adoption}
\label{subsubsec:sectoral_adoption}
To analyze deployment patterns across sectors, we first categorize domains in the stable panel using SafeDNS domain classification \cite{safeDns}. This taxonomy assigns domains to functional service categories based on the primary content or services they host.

From this categorization, we focus on sectors that commonly host services exchanging sensitive or high-value information with long confidentiality lifetimes including, government services, banking and financial transactions, healthcare-related services, personal communications (e.g., webmail) and platforms hosting private or regulated content.

Across these policy-relevant sectors, PQ-TLS adoption varies substantially. By March~2026, e-commerce $(n=15{,}506)$ and social networks $(n=1{,}525)$ exhibit the highest overall adoption, with approximately 68\% and 62\% of domains defaulting to PQ-TLS. Finance $(n=24{,}563)$ and healthcare-related services $(n=13{,}704)$ follow, with adoption levels of 59\% and 51\%, respectively. File storage $(n=316)$ and webmail $(n=568)$ show moderate uptake at around 46\% and 41\%. In contrast, government services $(n=18{,}178)$ and chats \& messengers $(n=768)$ lag behind, with adoption levels of 38\% and 27\%, respectively. These results indicate that, \textbf{even in policy-prioritized sectors that handle sensitive or long-lived data, PQ deployment remains partial and uneven} across the ecosystem.

While our analysis focuses on policy-relevant sectors, Appendix-\ref{appendix:sector_distribution} reports PQ-TLS adoption across all SafeDNS categories for completeness.

\subsubsection{National Adoption}
\label{subsubsec:national_adoption}

To characterize national and regional variation in PQ-TLS deployment, we analyze country-specific domain samples drawn from jurisdictions that have published national or regional PQC transition strategies. For each country or region listed in Table~\ref{tab:1_pqc_policies}, we construct domain samples based on country-code top-level domains (ccTLDs). Specifically, we scan the Top 10{,}000 nationally relevant domains under the corresponding ccTLD (e.g., \texttt{.au} for Australia) and the Top 1{,}000 government-operated domains under the associated government subdomain (e.g., \texttt{.gov.au}), yielding 11{,}000 targets per jurisdiction.

Across the general domain sample, PQ-TLS adoption differs markedly across countries. By March~2026, the USA and Australia exhibit the highest adoption levels, with approximately 57\% and 53\% of domains defaulting to PQ-TLS. The United Kingdom and India follow, with adoption levels around 45\% and 40\%. In contrast, France and Germany show substantially lower adoption, at 27\% and 16\%, respectively.

A similar but more pronounced pattern is observed among government-operated domains. By March~2026, Australia and the United Kingdom again exhibit the highest adoption levels, at 57\% and 47\%, respectively, although these values remain modest in absolute terms. \textbf{In countries such as Canada, Germany, and India, adoption remains extremely limited, with fewer than 7\% of government domains defaulting to PQ-TLS.}

\subsubsection{Infrastructure Mediation}
To understand the drivers of sectoral and national adoption patterns, we analyze PQ-TLS deployment through the lens of deployment agency (\S~\ref{subsec:tls_management_attribution}).

Figure~\ref{fig:2_policy_alignment} shows that PQ-TLS adoption across sectors, national top domains, and government services is strongly mediated by the underlying infrastructure. Across all three views, domains served via managed infrastructure platforms exhibit consistently high adoption rates, typically ranging from approximately 50\% to over 80\%, while owner-managed deployments remain markedly low, often below 10\% and, in many government contexts, below 3\%. These patterns indicate that \textbf{even in sectors prioritized by national policies, observable PQ-TLS adoption is concentrated among domains that inherit PQ-TLS through managed infrastructure providers}.

\subsubsection{Adoption Alignment}
Given this strong infrastructure influence, we examine whether national transition timelines correspond to observable differences in PQ-TLS adoption. However, comparing these observations with national transition timelines reveals no clear evidence that earlier policy targets correspond to higher observable PQ-TLS deployment. As Figure~\ref{fig:2_policy_alignment}-(b) shows, \textbf{jurisdictions with earlier transition targets (2030--2031) exhibit lower mean adoption rate (32.4\%) than those with later horizons (45.7\%)}, suggesting counterintuitive alignment with policy timelines.

This apparent reversal is largely explained by deployment composition. When restricting the analysis to potentially owner-managed domains, adoption remains uniformly low across both groups (9.6\% vs.\ 8.4\%). Likewise, within Cloudflare-hosted domains, which provide the largest common managed platform across jurisdictions, adoption rates are highly similar across countries (93.5\% vs.\ 94.1\%).

Taken together, these results indicate that \textbf{cross-sector and cross-jurisdiction variation in PQ-TLS adoption aligns more closely with managed infrastructure composition than with sectoral priorities or policy timelines}.

\begin{summary}
\noindent\textbf{Key Takeaways for RQ3:}
\textit{Observable PQ-TLS deployment shows limited alignment with the sectoral priorities and timelines outlined in national PQC policies. Across sectors and jurisdictions, PQ-enabled deployments are predominantly associated with domains served via managed infrastructure platforms, while owner-managed domains, including government-operated services, remain largely classical.}
\end{summary}
\vspace{-4pt}

\subsection{Operational Viability}
\label{subsec:rq5_operational}
To provide empirical feedback on the performance and compatibility implications of PQ key exchange, we examine the PQ-TLS deployment along three dimensions: handshake performance (\S~\ref{subsubsec:handshake_performance}), negotiation reliability (\S~\ref{subsubsec:failure_modes}), and geographic consistency (\S~\ref{subsubsec:geographic_consistency}).

\subsubsection{Handshake performance}
\label{subsubsec:handshake_performance}

\input{figures/3_latency/3_latency_figure}
We evaluate the performance impact of PQ key exchange using the differential measurement methodology described in \S~\ref{sec:methodology}, reporting averages over repeated handshakes to mitigate the network-induced variability. Our primary comparison focuses on the hybrid key exchange group \texttt{X25519MLKEM768} against its classical counterpart \texttt{X25519}, isolating the incremental cost of the PQ component.

Our measurements across 328{,}290 domains that successfully negotiated both configurations show that \textbf{PQ-TLS does not result in a meaningful increase in TLS handshake latency}. As shown in Figure~\ref{fig:3_latency}, the median delta is 0\,ms, with an interquartile range (IQR) of $[-5, +5]$\,ms. Nearly half of the domains (45.90\%) exhibit a negative delta, while an additional 31.26\% fall within 0–5\,ms. Only 13.02\% of domains exhibit a delta exceeding 10\,ms. These negative deltas may not necessarily imply that PQ key exchange is intrinsically faster; rather, they indicate that the overhead of the hybrid construction is effectively masked by normal network latency variation and connection setup noise.

Absolute handshake time distributions further support this observation. For classical \texttt{X25519}, the median handshake time is 32\,ms, with an IQR of $[26.0,61.0]$\,ms. Corresponding values for \texttt{X25519MLKEM768} are nearly identical, with the same median of 32\,ms and an IQR of $[26.0,59.0]$\,ms. Even at the tail, the 90th-percentile handshake times differ by only 5\,ms (456\,ms for \texttt{X25519} vs. 451\,ms for \texttt{X25519MLKEM768}), indicating that hybrid PQ key exchange does not materially increase end-to-end handshake latency. 

\begin{sloppypar}
While latency impact is negligible, \textbf{hybrid PQ key exchange introduces a predictable increase in handshake message size}. Relative to the classical baseline, \texttt{X25519MLKEM768} increases client-to-server traffic by a median of 1{,}176\,bytes and server-to-client traffic by 1{,}088\,bytes. This overhead is highly consistent across domains, reflecting the deterministic size of ML-KEM public keys.     
\end{sloppypar}

\subsubsection{Negotiation reliability and failure modes}
\label{subsubsec:failure_modes}

To assess whether increased handshake message sizes or altered negotiation structure introduce compatibility issues, we analyze TLS handshake failures across the full Top~1M domain set. Of the 315{,}506 domains that fail to complete a TLS handshake, 57.08\% fail before any cryptographic negotiation occurs: DNS resolution and reachability errors account for 105{,}189 domains (33.34\%), while TCP and transport-layer failures account for an additional 74{,}901 domains (23.74\%). These failures typically manifest as \emph{connection refusals}, \emph{unexpected EOF}, or write errors (e.g., \texttt{errno=104}).

A further 74{,}585 domains (23.64\%) explicitly reject TLS~1.3 connections due to protocol version or policy constraint related issues. As our client advertises only TLS~1.3 to enable PQ key exchange, these failures reflect a lack of support for contemporary TLS versions rather than incompatibility with PQC algorithms. 

Only 60{,}829 domains (19.28\%) fail during the TLS handshake phase where cryptographic parameters are negotiated. Even within this subset, failures are dominated by generic negotiation errors such as \emph{no common cipher suite}, which account for 49{,}089 cases (80.7\% of handshake-phase failures) and typically reflect restrictive classical configurations. Other handshake failures are comparatively rare and include SNI-related rejections (e.g., \emph{unrecognized name}), server internal errors, malformed message handling (e.g., decode errors), and packet-level violations. Collectively, these account for less than 4\% of handshake-phase failures.

While certain low-frequency errors, such as \emph{decode errors} or \emph{packet length violations}, could theoretically arise from increased handshake message sizes, their extremely limited occurrence (less than 0.16\% of handshake-phase failures), combined with the absence of correlated performance degradation, suggests that \textbf{MTU-related or fragmentation-induced failures are not a significant barrier to PQ key exchange deployment in current Internet practice}.

\subsubsection{Geographic consistency}
\label{subsubsec:geographic_consistency}

In the July~2025 measurement round, we observe 468 domains (0.05\% of the stable panel) that negotiate different default key exchange mechanisms depending on client location. For these domains, some regions expose hybrid PQ key exchange while others continue to negotiate classical groups.

Many of these domains are hosted on managed infrastructure platforms deployed via globally distributed CDNs (e.g., Amazon CloudFront), where configuration updates may propagate gradually across geographically distributed endpoints. As a result, different regions may reflect different stages of PQ-TLS deployment at a given point in time.

However, in subsequent measurement rounds, the number of domains exhibiting region-dependent negotiation decreases substantially, and by March~2026 most previously inconsistent domains converge to a uniform configuration across all vantage points. This convergence suggests that early PQ-TLS rollouts may initially appear regionally uneven due to gradual configuration propagation across distributed infrastructure, but tend to stabilize as deployments mature.

Across all measurement rounds, we do not observe regional differences in handshake latency, message size overhead, or failure characteristics once PQ-TLS is negotiated. These results indicate that \textbf{while early PQ deployments temporarily expose inconsistent configurations across regions, the operational behavior of PQ-TLS remains consistent once configurations are uniformly deployed}.

\begin{summary}
\noindent\textbf{Key Takeaways for RQ4:}
\textit{PQ-TLS deployment exhibits strong operational viability across performance, reliability, and consistency dimensions. At Internet scale, hybrid PQ key exchange introduces no meaningful increase in TLS handshake latency. Compatibility issues are largely unrelated to PQ mechanisms, instead arising from legacy protocol constraints and general network conditions. PQ-TLS behavior remains consistent across regions once uniformly deployed.}
\end{summary}

\subsection{Security Co-evolution}
\label{subsec:rq5_security_depth}
To examine whether PQ adoption coincides with broader improvements in protocol-level security posture, we compare PQ and classical TLS domains while explicitly accounting for deployment agency.

\begin{sloppypar}
Our analysis evaluates four dimensions of protocol-level security: support for legacy protocol versions (e.g., TLS~1.0, TLS~1.1), deprecated cipher suites (e.g., 3DES, IDEA, and obsoleted constructions), indicators associated with historically weak TLS configurations (e.g., \textsc{BEAST}, \textsc{BREACH}, \textsc{Lucky13}, \textsc{Sweet32}, and RC4 support), and certificate configuration issues (e.g., hostname mismatch and chain-of-trust errors). Full labeling criteria are provided in Appendix~\ref{appendix:security_depth}. 
\end{sloppypar}

To minimize operational impact, we restrict measurements to non-intrusive tests and enforce a delay of five minutes between successive scans of the same domain, following \cite{menlo_report}.

\subsubsection{Provider-managed domains.}
For \emph{potentially infrastructure provider managed} domains, we compare PQ-TLS and classical TLS within the same providers, retaining those with at least 1{,}000 domains in each class and sampling evenly.

Under this provider-balanced comparison, \textbf{PQ-TLS domains more frequently retain legacy compatibility features}. Approximately 18.0\% support TLS~1.0/TLS~1.1 compared to 14.7\% of matched classical domains, and 73.6\% advertise at least one deprecated cipher suite versus 49.9\%. This pattern extends to vulnerability indicators: PQ-TLS domains more frequently retain legacy cryptographic features, such as CBC-mode ciphers, TLS~1.0-era record handling, and compression mechanisms, which corresponds to a higher prevalence of indicators associated with \textsc{BEAST} (17.2\% vs.\ 14.3\%), \textsc{BREACH} (50.6\% vs.\ 31.7\%), while support for stream ciphers such as, RC4 is absent in both classes.

\textbf{Certificate-management signals are stronger among PQ-TLS deployments}. OCSP stapling is more common (18.2\% vs.\ 3.9\%), hostname mismatches are less frequent (0.7\% vs.\ 10.4\%), and chain-of-trust issues are rare in both groups.

\begin{sloppypar}
These patterns vary across providers, reflecting platform-specific configurations. For example, Cloudflare exhibits higher legacy protocol and cipher support among PQ-TLS domains, Fastly shows lower legacy protocol support but higher deprecated cipher usage, while Amazon CloudFront presents a mixed profile. These differences indicate that \textbf{managed PQ-TLS deployments largely inherit the TLS compatibility profile of the underlying platform}.
\end{sloppypar}

\subsubsection{Owner-managed domains.}
For domains classified as \emph{potentially owner managed}, we compare randomly sample PQ-TLS and classical TLS domains (10{,}000 each). 

In contrast, \textbf{PQ-TLS adoption among owner-managed domains is associated with modestly stricter configurations}, with lower support for legacy protocol versions (10.5\% vs.\ 11.9\%) and deprecated cipher suites (46.9\% vs.\ 52.6\%). This trend is also reflected in a lower prevalence of vulnerability indicators associated with \textsc{BEAST} (8.0\% vs.\ 10.0\%), \textsc{BREACH} (40.0\% vs.\ 43.1\%), and \textsc{Lucky13} (46.6\% vs.\ 52.3\%), along with reduced support for deprecated stream ciphers such as RC4 (0.04\% vs.\ 0.30\%), although indicators linked to 64-bit block cipher usage (\textsc{Sweet32}) remain more common (6.8\% vs.\ 4.5\%).

Certificate-related signals align similarly, with higher OCSP stapling rates (12.9\% vs.\ 6.3\%), fewer hostname mismatches (5.0\% vs.\ 7.5\%) and chain-of-trust issues (2.9\% vs.\ 5.9\%). 

Taken together, these results indicate that \textbf{security co-evolution during early PQ-TLS deployment is strongly shaped by \emph{deployment agency}}. Among provider-managed domains, PQ-TLS is typically introduced through platforms that combine strong certificate automation with broad backward compatibility. Among owner-managed domains, PQ-TLS adoption more closely reflects deliberate protocol modernization.

\begin{summary}
\noindent\textbf{Key Takeaways for RQ5:}
\textit{PQ-TLS adoption does not imply a uniform improvement in protocol-level security. Provider-managed deployments inherit platform-level configurations that combine stronger certificate practices with greater retention of legacy protocol features, while owner-managed deployments are associated with modestly stricter TLS configurations.}
\end{summary}

%% file: tables/0_pqc_adoption.tex
\begin{table}[!t]
\renewcommand{\arraystretch}{1}
    \caption{Default negotiation preference versus advertised support for PQ key exchange (n = 684{,}494)}
    \label{tab:0_pqc_adoption}
    
    \centering
    \resizebox{\columnwidth}{!}{
        \rowcolors{2}{}{lightgray}        
        \begin{tabular}{*{1}{L{3.2cm}}*{6}{M{1.2cm}}}
        
            \toprule
                
                \multicolumn{1}{c}{\multirow{2}{*}{\textbf{PQC KEM}}} & 
                \multicolumn{3}{c}{\textbf{Default}} &
                \multicolumn{3}{c}{\textbf{Support}} \\ 
                \cmidrule(lr){2-4} \cmidrule(lr){5-7}
                \rowcolor{white}
                &
                \textbf{Jul-25} &
                \textbf{Nov-25} &
                \textbf{Mar-26} &
                
                \textbf{Jul-25} &
                \textbf{Nov-25} &
                \textbf{Mar-26} \\
            \midrule
                \multicolumn{5}{l}{\textit{Hybrid KEMs}} \\
                X25519MLKEM768     & 31.26\% (214,020)  & 47.37\% (324,276)  & 49.22\% (336,919) &  31.27\% (214,093) & 47.37\% (324,288)   & 49.22\% (336,921) \\
                SecP256r1MLKEM768  & 0 & 0              & 0                  & 1.02\% (6,981)    &  3.64\% (24,937)                         & 5.53\% (37,831)  \\
                SecP384r1MLKEM1024 & 0 & 0              & 0                  & 0.01\% (83)       &  0.07\% (477)                            & 1.95\% (13,348) \\
            \midrule
                
                \multicolumn{5}{l}{\textit{Pure PQC KEMs}} \\
                ML-KEM-512         & 0 & 0       & 0                         & 0 & $\approx0.00\%$ (2)       & $\approx0.00\%$ (3)  \\
                ML-KEM-768         & 0 & 0       & 0                         & 0 & $\approx0.00\%$ (3)       & $\approx0.00\%$ (6)  \\
                ML-KEM-1024        & 0 & 0       & 0                         & 0 & $\approx0.00\%$ (3)       & 0.66\% (4,592)  \\

            \bottomrule
            \addlinespace[5pt]
            \multicolumn{7}{l}{ 
                \begin{tabular}[l]{@{}l@{}}
                    \rowcolor{white}
                    \normalsize\textbf{Default}: Indicates the KEM selected during an unconstrained TLS~1.3 handshake. \\
                    \addlinespace[5pt]
                    \normalsize\textbf{Support}: Indicates KEMs that successfully negotiated when explicitly requested. 
                    
                \end{tabular}
                
            }
            
        \end{tabular}
    }
\end{table}

%% file: figures/1_infra_centralisation/1_1_config_attribution_figure.tex
% Wrapper name differs from the PDF basename so arXiv preserves the figure asset.
\begin{figure}[!t]
    \centering
    \includegraphics[width=\columnwidth]{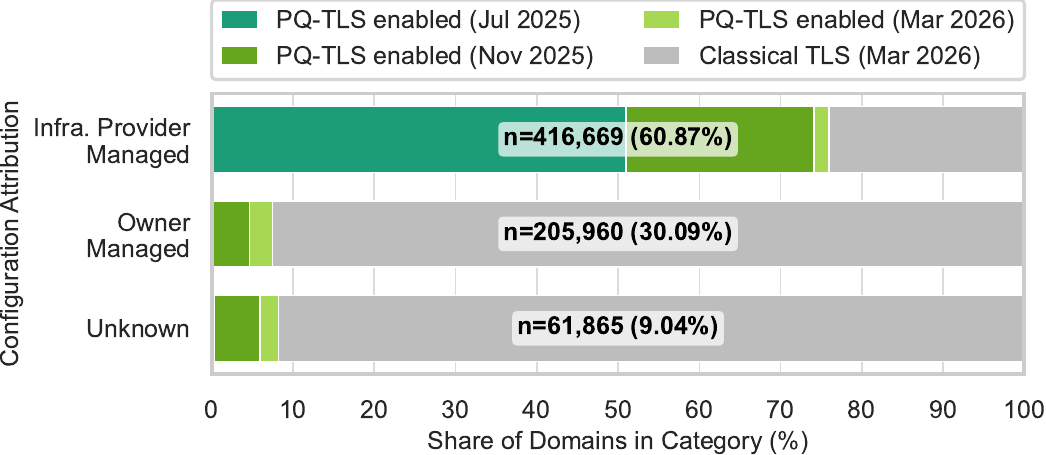}
    
    \caption{PQ-TLS adoption by management attribution.}

    \label{fig:1_1_config_attribution}    
    
\end{figure}

%% file: figures/1_infra_centralisation/2_infra_providers.tex
\begin{figure*}[!t]
    \centering
    
    \subfloat[
        \centering Distribution of domains defaulting to PQ-TLS across infrastructure providers.]{%
        \raisebox{0.1\height}{
            \includegraphics[width=0.96\columnwidth]{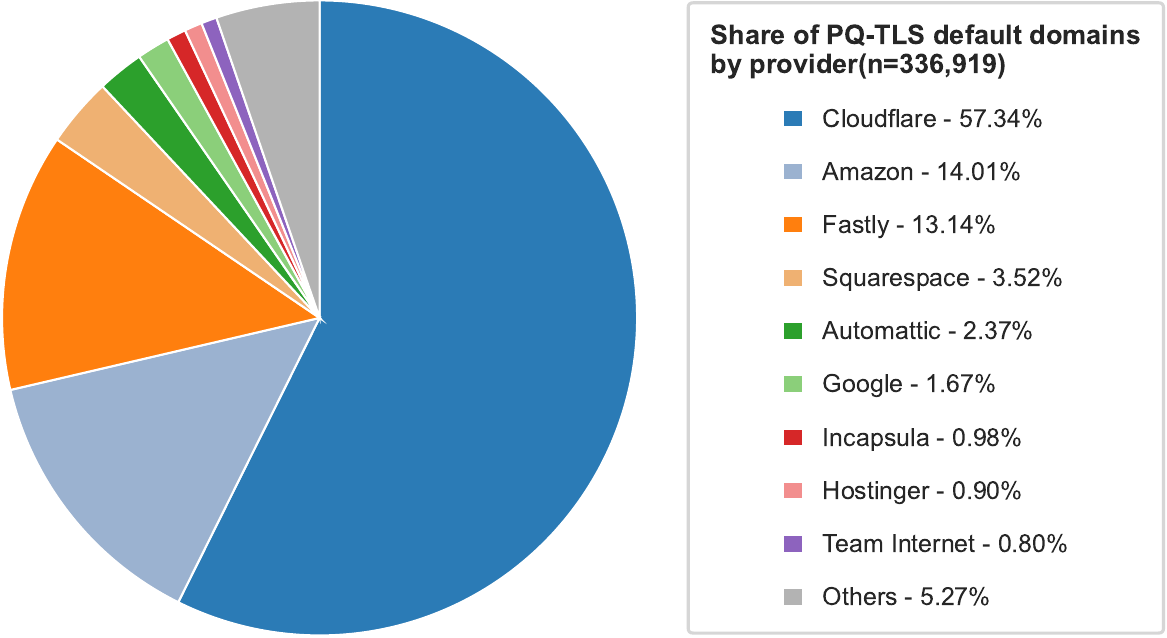}
        }%
        \label{fig:2_a_top_pqc_providers}
    }
    \hfil
    \hspace{1.2em}
    \subfloat[
        \centering PQC adoption across the infrastructure providers hosting most PQ-TLS domains.]{%
        \includegraphics[width=0.96\columnwidth]
        {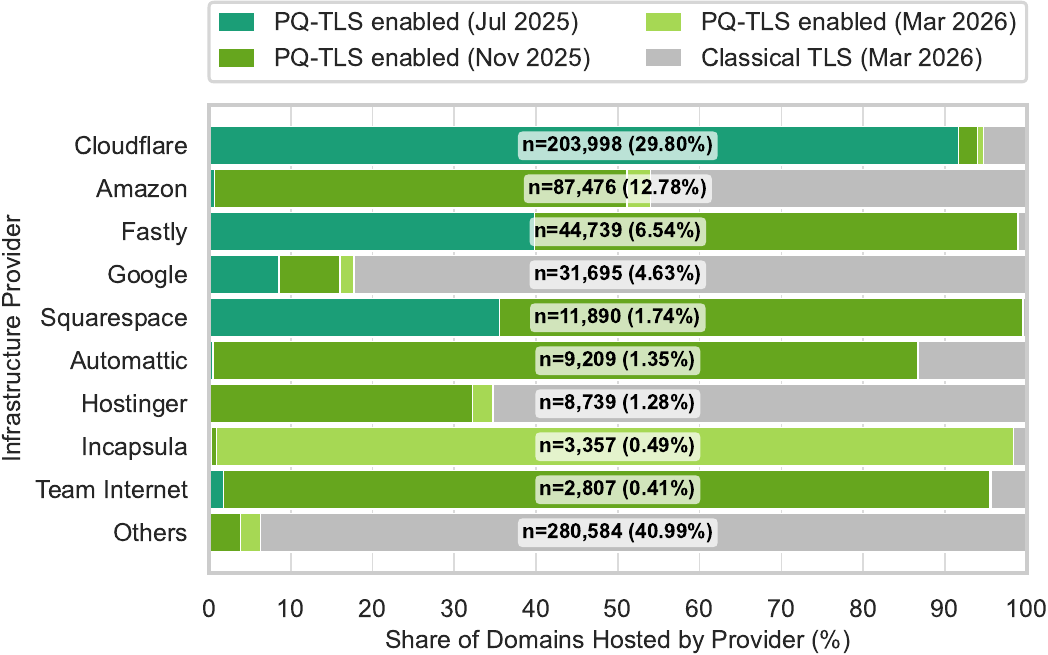}%
        \label{fig:2_b_top_pqc_providers}%
    }
    \vspace{1em}
    \justifying\noindent Each bar in Subfigure (b) is annotated with the absolute number of served domains and the corresponding fraction of all successful scans ($n = 684{,}494$).
    
    \caption{Infrastructure concentration of PQ-TLS deployment by March 2026.}
    \label{fig:2_infrastructure_providers}
    
\end{figure*}

%% file: figures/2_sectors_and_countries/2_policy_alignment_figure.tex
% Wrapper name differs from the PDF basename so arXiv preserves the figure asset.
\begin{figure*}[!t]
    \centering
    \includegraphics[width=\textwidth]{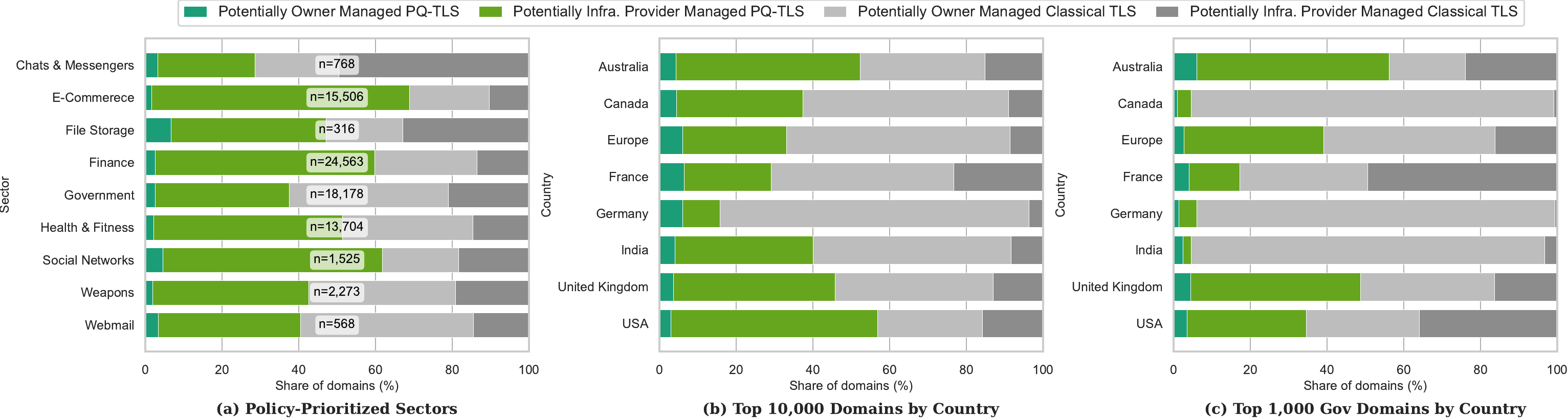}
    
    \caption{PQ-TLS adoption across sensitive sectors and Jurisdictions (March 2026)}
    \label{fig:2_policy_alignment}    
\end{figure*}

%% file: figures/3_latency/3_latency_figure.tex
% Wrapper name differs from the PDF basename so arXiv preserves the figure asset.
\begin{figure*}[!t]
    \centering
    \includegraphics[width=0.85\textwidth]{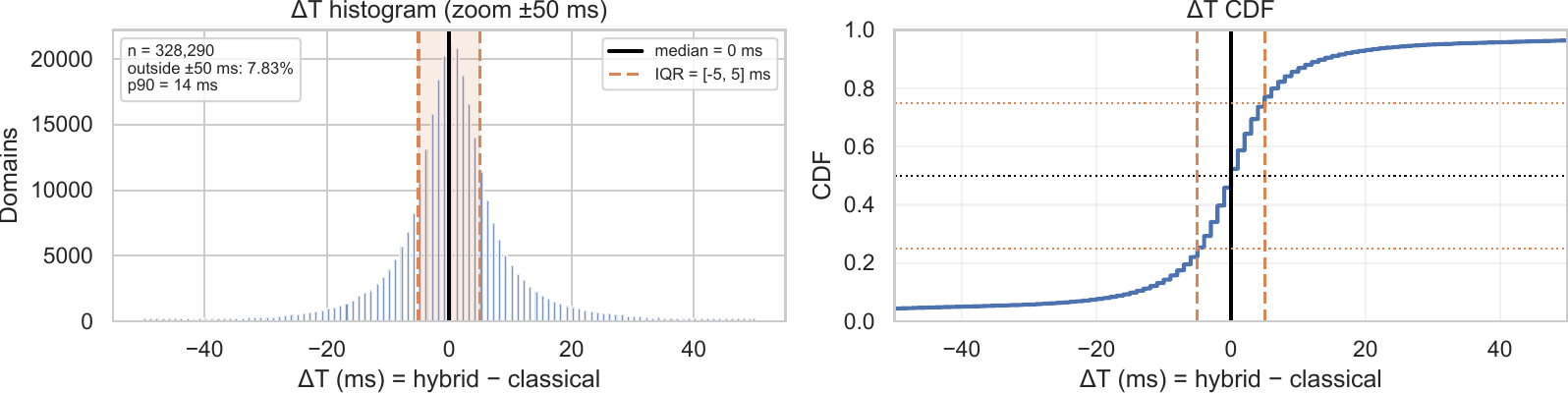}

    \textbf{Left}: Histogram of $\Delta_T$ zoomed to $\pm50$\,ms, showing strong concentration around 0\,ms with median 0\,ms and IQR $[-5,5]$\,ms. \\ \textbf{Right}: Empirical CDF of $\Delta_T$, indicating that over 90\% of domains experience a latency difference of at most 14\,ms
    
    \caption{Per-domain handshake latency difference $\Delta_T = T_{\text{hybrid}} - T_{\text{classical}}$ for \texttt{X25519MLKEM768} relative to \texttt{X25519}.}

    \label{fig:3_latency}    
    
\end{figure*}

%% file: sections/7_discussion.tex
\section{Discussion}
\label{sec:discussion}
\enlargethispage{\baselineskip}
This section synthesizes the key findings from our measurements and examines their implications for ongoing PQC transition efforts. By contrasting policy expectations with observed deployment behavior, we highlight where current strategies align with operational realities and where adjustments may be necessary to better support the transition.

\textbf{Configuration Convergence in Practice.}  
Although national PQC transition policies frequently emphasize algorithm agility and recommend multiple candidate constructions, our measurements show that operational TLS deployments have effectively converged on a single hybrid configuration: \texttt{X25519MLKEM768}. Alternative constructions appear only rarely as secondary capabilities and are almost never selected as default negotiation choices. This convergence reflects practical deployment constraints, including interoperability requirements, browser defaults, and infrastructure-provider support, which naturally steer the ecosystem toward a small set of widely supported configurations. Policy guidance that emphasizes broad algorithmic diversity without corresponding deployment incentives risks becoming more aspirational than actionable. Policymakers may therefore achieve greater impact by recognizing de facto convergence and focusing guidance on the deployment, monitoring, and evolution of dominant constructions while keeping the recommended design space minimal and interoperable.

\textbf{Infrastructure Providers as Primary Deployment Actors.}  
Our analysis shows that the majority of observable PQ-TLS deployment is driven by a small number of managed infrastructure providers rather than independent migration decisions by individual domain operators. Content delivery networks and cloud platforms effectively act as deployment multipliers: once a provider enables PQ-TLS, thousands of served domains inherit the capability simultaneously. This architecture significantly accelerates visible adoption and also concentrates influence over the transition within a small number of actors. Engaging infrastructure providers directly through guidance, coordination, and transparency mechanisms may therefore offer greater leverage, given their central role in shaping deployment across the Internet.

\textbf{Understanding National Adoption Differences.}  
Our sectoral analysis indicates that PQ-TLS deployment does not uniformly follow the sectoral priorities articulated in many national transition strategies. Sectors commonly highlighted in policy discussions, such as government, exhibit comparatively low levels of adoption, while several platform-driven consumer services show substantially higher deployment. This disparity reflects structural differences in service provisioning: many consumer-facing platforms inherit cryptographic capabilities through shared managed infrastructure, whereas sectors such as government and healthcare face procurement cycles, legacy system dependencies, and regulatory constraints that slow the adoption of new cryptographic mechanisms \cite{crypto_deployment_challenges}. These findings suggest that policies may benefit from distinguishing between sectors that can passively inherit security properties through shared infrastructure and those that require targeted support, funding, or technical assistance to enable migration within more constrained operational environments.

\textbf{Operational Feasibility of PQ-TLS.}  
Concerns about performance, compatibility, and reliability often appear implicitly in PQC transition discussions, yet our measurements show that PQ-TLS introduces no meaningful latency penalty and causes no systemic negotiation failures at Internet scale. Observed failures are dominated by pre-existing DNS, transport, and TLS version issues rather than PQC itself. These results support the technical feasibility assumptions underlying many policies, while underscoring the value of empirical validation. Embedding continuous measurement into transition planning can help policymakers distinguish genuine from perceived deployment risks and adjust their guidance accordingly as the ecosystem evolves.

\textbf{PQC Readiness and Broader TLS Security Practices.}  
Our analysis shows that PQ-TLS adoption does not correspond to a uniform shift toward stricter TLS configurations; instead, its security implications depend strongly on deployment context. Provider-managed domains show strong certificate automation with continued support for legacy protocol versions and deprecated cipher suites, reflecting a compatibility-oriented deployment model. In contrast, PQ-TLS adoption among owner-managed domains is associated with modestly stricter configurations, including reduced support for legacy features. These patterns indicate that PQC transition is not a simple cryptographic substitution, but is shaped by how and where deployment occurs. More broadly, they highlight that effective security outcomes depend not only on adopting new cryptographic mechanisms, but also on the surrounding protocol configuration and operational practices in which they are embedded.

\textbf{Implications for PQC Transition Efforts.}  
Taken together, these findings suggest that the PQC transition is currently progressing primarily through infrastructure-mediated deployment rather than through coordinated migration by independently managed systems. While large CDNs and cloud platforms can rapidly introduce PQ-TLS across thousands of hosted domains, most owner-managed services depend on operating system distributions and packaged software stacks for cryptographic capabilities. Support for PQ-enabled TLS has already been incorporated into widely used cryptographic libraries and server software for roughly a year. For example, OpenSSL~3.5.0, released in April~2025, introduced built-in support for the NIST-standardized PQC algorithms \cite{openssllibraryOpenSSLFinal}. Yet our measurements show that adoption outside large infrastructure platforms remains limited. This gap indicates that software availability alone is insufficient to drive deployment. Supporting the broader transition will therefore require more than technical standardization of \textit{which} algorithms to adopt by \textit{when}. Practical guidance, deployment tooling, and operator education will be important to help organizations understand \textit{how} to enable PQ mechanisms within their infrastructure and integrate them safely into existing operational environments.

\textbf{Acceleration and Limits of Security Transitions.}  
Historical transitions in TLS illustrate both the accelerating pace and the limits of security adoption at Internet scale. The shift from TLS~1.1 to TLS~1.2 required nearly five years to reach 15\% deployment, reflecting a largely operator-driven upgrade process. In contrast, driven by a small number of major infrastructure providers, TLS~1.3 reached nearly 48\% adoption within two years and four months following its standardization in August~2018 \cite{www-21-tls-1.3-audit}. We observe a similar but even more pronounced trend for PQ-TLS, which achieves comparable adoption levels (47.37\%) in approximately one year and three months (November~2025), driven by the same infrastructure-mediated deployment model.
However, accelerated early adoption does not necessarily imply complete transition. Even more than seven years after standardization, TLS~1.3 has not yet reached universal deployment (\S~\ref{subsubsec:failure_modes}), reflecting persistent long-tail inertia among systems. For PQC, this long tail represents a growing security risk. Successful PQC transition will therefore require sustained effort to mitigate inertia through early preparation, stronger coordination across stakeholders, and continued investment to support secure and widespread deployment.

%% file: sections/8_limitations.tex
\section{Limitations}
\label{subsec:limitations}
Our measurements rely on active scanning of public-facing domains drawn from the Top~1M, and therefore capture only the \emph{public Internet}. We do not observe PQ deployments within private intranets, VPNs, or other internal networks not exposed to the public Internet. While some national PQC strategies prioritize national or classified systems (e.g., CNSA 2.0 \cite{us_2}), many published transition roadmaps (e.g., BSI \cite{germany_2}) address a broader set of systems, including Internet-facing public services. Measuring the public web therefore remains the most appropriate proxy for evaluating the externally observable component of these policy mandates.

Our measurements focus exclusively on TLS as used for HTTPS. While TLS is also employed by other application protocols (e.g., email, messaging, and custom services), HTTPS dominates Internet-facing cryptographic traffic \cite{measuring_https_adoption}. As such, HTTPS-based TLS provides the most operationally significant surface for evaluating PQ deployment at Internet scale.

All measurements were conducted from cloud-based vantage points using distributed measurement infrastructure. While this setup enables globally comparable observations, it may not fully capture the experience of end users operating from residential networks, enterprise environments, or constrained network conditions. Factors such as last-mile latency, network congestion, and device heterogeneity can influence handshake performance and reliability in ways not reflected in our study. As a result, observed operational viability should be interpreted as representative of well-provisioned network paths rather than all end-user environments.

Taken together, these limitations define the intended scope of our results. We empirically assess PQ-TLS readiness, configuration, and operational viability on the public web, in line with contemporary PQC transition policies.

%% file: sections/9_conclusion.tex
\section{Conclusion}
\label{sec:conclusion}

This paper presents the first Internet-scale longitudinal measurement study of PQ-TLS deployment. By analyzing more than 2 billion TLS handshakes across one million domains, collected from 11 globally distributed vantage points, we provide an empirical view of how PQC is transitioning from standardization into real-world deployment.

Our measurements reveal a persistent gap between national policy expectations and deployment reality. While nearly half of the stable domain panel defaults to PQ key exchange by March~2026, deployment is narrowly concentrated around a single hybrid construction, \texttt{X25519MLKEM768}. PQ-TLS adoption is overwhelmingly driven by managed infrastructure providers, while owner-managed and government-operated domains largely default to classical. Across sectors and countries, observable deployment shows limited correspondence with policy timelines and priorities.

At the same time, our results demonstrate that PQ-TLS is operationally viable at Internet scale. Hybrid key exchange introduces no meaningful increase in handshake latency, and PQ mechanisms do not result in observable compatibility failures. However, PQ adoption does not necessarily coincide with broader TLS hardening, as domains often retain legacy protocol versions and deprecated cipher support.

These findings suggest that successful PQC transition will require more than algorithm standardization and calendar-based migration targets. Policies should account for the central role of infrastructure providers, support owner-managed systems with concrete deployment guidance and tooling, and incorporate empirical protocol-level measurements into progress tracking. Without this operational focus, PQC migration may become visible in aggregate statistics while remaining uneven across sectors, dependent on a small number of platforms, and only partially aligned with the security outcomes transition policies aim to achieve.

%% file: sections/references.tex
\bibliographystyle{ACM-Reference-Format}
% \balance
\bibliography{references}

%% file: sections/ethics.tex
\section*{Ethics}
\label{sec:ethics}
Internet-wide scanning is a well-established practice to understand the topology and security posture of the global network \cite{predicting_ipv4_services, iwv_iws}, with prior work demonstrating its importance for assessing systemic risks (see \S\ref{sec:literature_review}). Our study adheres to this tradition, employing only lightweight TLS handshakes, which we gracefully terminate immediately after handshake completion and before any application-layer data exchange. They do not attempt to exploit vulnerabilities, log in, or access protected resources. To further minimize potential impact, we instrumented our tools to enforce a minimum delay of five minutes between successive handshakes to the same domain. Prior research has shown that exchanging TLS handshake messages is trivial \cite{www-21-tls-1.3-audit} and that the vast majority of network operators do not regard such scanning as a significant threat \cite{iwv_iws}.

We also followed community standards for ethical network measurement. Scans were conducted from dedicated IP addresses with Reverse DNS records pointing to a project webpage that described the study's purpose and provided an opt-out mechanism allowing operators to exclude their infrastructure from future scans. Our approach builds on established practices for minimizing disruption to Internet services and aligns with ethical frameworks for measurement research, such as the Menlo Report \cite{menlo_report} and guidelines on good Internet citizenship \cite{zmap}.

Our methodology was reviewed by the Human Ethics Board of our institution, which confirmed that no personally identifiable information (PII) is collected and therefore no human subjects approval was required. All identifiable entities in our released datasets are irreversibly hashed.

Finally, we emphasize the public-interest motivation of this work. TLS underpins the confidentiality and integrity of modern Internet communication, and understanding its preparedness for post-quantum transition has clear societal and security relevance. We therefore believe that the minimal risks associated with our non-invasive measurements are outweighed by the importance of the insights shared in this study.

%% file: sections/appendix.tex
\newpage
\appendix

\input{sections/2_background}

\section{Policy Collection and Inclusion Criteria}
\label{appendix:shortlisted_countries}

We identify countries with demonstrated engagement in PQC, reflecting early technical investment and institutional preparedness for transition. Specifically, we include all countries that contributed submissions to NIST’s Round~1 PQC standardization process, which comprised 82 proposals from 25 countries~\cite{nist_first_round,ship_has_sailed}. 

\input{tables/2_shortlisted_countries} 

We further expand this set by incorporating the world’s largest economic blocs, as identified by World Bank GDP rankings~\cite{world_bank}, under the rationale that successful PQC adoption within these regions is likely to exert disproportionate global influence through market scale and infrastructure deployment. 

Together, these criteria yield a shortlist of 31 countries, shown in Table~\ref{tab:2_shortlisted_countries}.

\subsection{Inclusion Criteria}

For each shortlisted country, we conduct web-based searches to identify PQC transition documents using structured keyword queries. These queries combine the \textit{country name} with terms such as “post-quantum cryptography”, “post-quantum cryptographic migration”, “post-quantum cryptographic transition”, and “post-quantum cryptography roadmap”. From the results, we prioritize primary sources, including government publications, cybersecurity agency guidance, and national standards organizations.

Policies are included if they satisfy the following criteria:
\begin{itemize}
    \item Explicitly reference NIST-standardized PQC algorithms and emphasize interoperability;
    \item Define sectoral scope or deployment priorities; and
    \item Articulate milestones or timelines for PQC migration or classical cryptography deprecation.
\end{itemize}

\section{PQ-TLS Adoption Across Tranco \& CrUX}
\label{appendix:tranco_crux}

To assess the robustness of our findings with respect to domain ranking sources, we replicate our measurements using two alternative widely used datasets, Tranco and Google CrUX. Table~\ref{tab:3_domcop_traco_crux_summary} and Table~\ref{tab:4_domcop_tranco_crux_adoption} summarize the scanning outcomes and resulting PQ-TLS adoption measurements across these domain lists.

Across all three datasets, PQ-TLS adoption rates and negotiated configurations remain highly consistent, confirming that our findings are not sensitive to the choice of domain ranking source.

\input{tables/3_domcop_traco_crux_summary}

\input{tables/4_domcop_tranco_crux_adoption}

\section{Sectoral Distribution of PQ-TLS Adoption}
\label{appendix:sector_distribution}

\input{figures/4_all_sectors/4_all_sectors_figure}

To provide a complete view of PQ-TLS deployment across the dataset, we examine adoption across all SafeDNS domain categories. SafeDNS classifies domains according to the primary type of content or service they host, enabling a broad characterization of sectoral deployment patterns beyond the policy-relevant sectors discussed in the main text.

Figure~\ref{fig:4_all_sectors} presents PQ-TLS default adoption rates across all SafeDNS categories observed in the stable panel for the July~2025, November~2025, and March~2026 measurement rounds. Each column corresponds to a category, with the sample size of domains in that category indicated in parentheses. Colors represent the percentage of domains within each category that negotiate PQ-TLS by default.

\textit{Note:} The domain counts shown in Figure~\ref{fig:4_all_sectors} include all domains assigned to each SafeDNS category. In contrast, the sectoral analysis presented in Section~\ref{subsec:rq3_policy_alignment} considers only domains for which TLS management attribution could be determined, excluding those labeled as \emph{Unknown}. Consequently, category totals may appear marginally larger in this figure.

\section{Protocol-level Security Depth: Labeling Criteria}
\label{appendix:security_depth}

This appendix documents the labeling criteria and test semantics used in our protocol-level security depth analysis (RQ4). Our goal is to characterize the \emph{configuration surface} exposed by operational TLS deployments rather than to assess exploitability or conduct adversarial testing.

All measurements were restricted to non-intrusive probes that do not modify server state, attempt credential extraction, or trigger repeated handshake failures. To further minimize operational impact, we enforce a minimum delay of five minutes between successive scans of the same domain.

\subsection{Protocol Version Support}
A domain is labeled as supporting an obsolete protocol version if it \emph{offers} any of the following during protocol negotiation, regardless of default preference:
\begin{itemize}
    \item SSLv2
    \item SSLv3
    \item TLS~1.0
    \item TLS~1.1
\end{itemize}
Support is detected through passive protocol negotiation and version advertisement. Domains that prefer TLS~1.3 but retain support for older versions are still classified as offering obsolete protocols, as such configurations may preserve downgrade paths.

\subsection{Cipher Suite Classification}
Cipher-related labels are assigned based on advertised cipher suite support according to OpenSSL\footnote{\url{https://docs.openssl.org/master/man1/openssl-ciphers/}} categorisation. A domain is flagged if it offers any cipher suites falling into the following categories:
\begin{itemize}
    \item \textbf{NULL Ciphers}: Provide no encryption.
    \item \textbf{aNULL Ciphers}: Provide encryption but no authentication, enabling man-in-the-middle attacks.
    \item \textbf{Export Ciphers}: Suites with intentionally weakened 40–56 bit keys, introduced for export control compliance and now trivially broken.
    \item \textbf{Low Ciphers}: Use outdated 64-bit block ciphers.
    \item \textbf{Obsoleted Ciphers}: Suites formally deprecated by standards bodies or removed from modern TLS protocols.
    \item \textbf{3DES/IDEA Ciphers}: Legacy block ciphers with outdated constructions or limited block sizes.
    \item \textbf{Strong-NoFS Ciphers}: Suites that employ strong encryption but lack forward secrecy, exposing past sessions if long-term keys are ever compromised.
\end{itemize}
Cipher availability is evaluated independently of negotiation preference.

\subsection{Known TLS Vulnerability Indicators}
We flag domains for \emph{potential exposure indicators} associated with well-known TLS weaknesses. These labels do not imply exploitability, only that the relevant protocol features are present. Indicators include:
\begin{itemize}
    \item \textsc{POODLE}: SSLv3 with CBC cipher support.
    \item \textsc{SWEET32}: support for 64-bit block ciphers.
    \item \textsc{FREAK}: EXPORT-grade RSA cipher support.
    \item \textsc{BEAST}: TLS~1.0 with CBC ciphers (even if higher versions are also supported).
    \item \textsc{Lucky13}: support for CBC-mode ciphers under TLS.
    \item \textsc{RC4}: RC4 cipher support.
    \item \textsc{LOGJAM}: DH EXPORT cipher support.
    \item \textsc{BREACH}: HTTP compression enabled, detected via advertised content-encoding headers on a single benign request to the root path.
\end{itemize}

All vulnerability indicators are evaluated conservatively and without active exploitation attempts. For compression-based indicators (e.g., \textsc{BREACH}), only passive detection of compression support is performed.

\subsection{Certificate Properties}
Certificate-related issues are identified using the following criteria:
\begin{itemize}
    \item \textbf{Certificate expiration}: the leaf certificate is expired at scan time.
    \item \textbf{Hostname mismatch}: the certificate does not correctly match the supplied domain name, including cases where matching relies solely on CN or wildcard behavior.
    \item \textbf{Chain of trust issues}: including incomplete chains, expired intermediates, self-signed certificates, or self-signed certificate authorities within the chain.
\end{itemize}

\subsection{OCSP Stapling}
OCSP stapling is evaluated based on server behavior during the TLS handshake. Domains are classified as not offering OCSP stapling if:
\begin{itemize}
    \item No stapled OCSP response is provided, or
    \item The stapled response contains an OCSP error (e.g., \texttt{unauthorized}).
\end{itemize}

\subsection{Interpretation}
Labels reflect the presence of legacy or potentially risky protocol features, not confirmed vulnerabilities. Many flagged conditions coexist with TLS~1.3 and modern cipher preferences, indicating residual compatibility rather than misconfiguration specific to PQC. As such, results should be interpreted as indicators of protocol hygiene rather than security failures.

%% file: sections/2_background.tex
\section{Cryptographic Background on PQ-TLS}
\label{appendix:background}

This section summarizes the cryptographic foundations of TLS, the quantum threat to its public-key components, and the standardization efforts that motivate the transition to PQC.

\subsection{Public-Key Cryptography in TLS}
Transport Layer Security (TLS) combines public-key and symmetric-key cryptography to provide authentication, integrity, and confidentiality for Internet communication. Public-key cryptography is used during the handshake for key establishment and authentication, while symmetric cryptography protects application data once a secure channel is established. Consequently, the security of the resulting session keys depends directly on the public-key mechanisms used during the handshake.

In TLS, key establishment is performed using ephemeral Diffie–Hellman (DH) exchanges. In practice, this is predominantly realized using elliptic-curve Diffie–Hellman (ECDH), most commonly over X25519 and, less frequently, X448. The exchange produces a shared secret from which symmetric session keys are derived. The use of ephemeral keys provides forward secrecy under the assumption that the underlying elliptic-curve problem remains hard \cite{rfc8446}.

Authentication in TLS is provided through digital signatures embedded in X.509 certificates. Today, the dominant signature algorithms are ECDSA over secp256r1 and secp384r1, and RSA using probabilistic signature schemes such as RSA-PSS. These signatures allow clients to authenticate servers and verify possession of the corresponding private keys \cite{rfc8446}.

\subsection{CRQC and Public-Key Risk}
A CRQC is a quantum system capable of executing Shor’s algorithm \cite{shors_algo} at a scale sufficient to break widely deployed public-key cryptographic schemes. Such a capability would compromise both ECDH-based key exchange (e.g., X25519) and classical digital signature algorithms (e.g., ECDSA and RSA).

This threat enables two classes of attacks against TLS. First, an adversary may record encrypted TLS traffic and later decrypt it by recovering session keys once the key establishment mechanism is broken, commonly referred to as a \textit{Harvest Now, Decrypt Later (HNDL)} attack. Second, an adversary may forge digital signatures, enabling server impersonation or man-in-the-middle attacks during connection establishment \cite{mosca_pqc}. 

By contrast, symmetric cryptography and hash functions are not directly broken by Shor’s algorithm and generally require only parameter changes. As a result, the quantum threat to TLS is concentrated on the public-key primitives used during the handshake.

\subsection{NIST’s PQC Standardization}
To address these risks, in August 2024, NIST finalized its first set of PQC standards, publishing FIPS 203 \cite{fips_203} for key encapsulation, FIPS 204 \cite{fips_204} for digital signatures, and FIPS 205 \cite{fips_205} for stateless hash-based signatures.

FIPS 203 standardizes ML-KEM, a lattice-based key encapsulation mechanism intended to serve as a PQ alternative to the key agreement functionality of classical Diffie–Hellman. FIPS 204 and FIPS 205 standardize ML-DSA and SLH-DSA for authentication. Together, these standards define PQ replacements for public-key primitives used by protocols such as TLS.

\subsection{International PQC Standardization Efforts}
Although NIST’s PQC standards form the primary interoperability baseline for Internet protocols, several national and regional standardization bodies have also proposed or evaluated alternative PQ algorithms. For example, the German Federal Office for Information Security (BSI) has introduced key establishment mechanisms such as FrodoKEM and Classic McEliece \cite{germany_2}. Similarly, the State Cryptography Administration (SCA) of China has supported domestic candidates including LAC, a lattice-based key establishment mechanism, as well as signature schemes such as AAGL and PQC-Sign \cite{china}. In South Korea, the Telecommunications Technology Association (TTA) has proposed SOLMAE \cite{sk}, a lattice-based key establishment mechanism designed for PQ key exchange.

Despite these parallel efforts, the algorithms standardized by NIST have emerged as the dominant global reference point for PQC deployment. NIST-selected primitives are currently being incorporated into protocol specifications developed within the Internet Engineering Task Force (IETF) \cite{ietf-uta-pqc-app-00} and are widely supported by major infrastructure providers (e.g., Cloudflare \cite{cloudflareDataExplorer}, Google \cite{googlePostquantumCryptography}) and cryptographic libraries \cite{pqc_client_side_survey}. In addition, international standards organizations such as ISO/IEC have initiated standardization activities based on NIST’s selected algorithms \cite{iso_sc27}. As a result, NIST-standardized PQC primitives effectively serve as the interoperability foundation for early PQC deployment in widely deployed Internet protocols such as TLS.

\subsection{PQC Deployment Urgency}
\label{subapp:pqc_urgency}
Although both key establishment and authentication are vulnerable to quantum attacks, their mitigation differs in urgency and deployment feasibility. Compromise of key establishment enables retrospective decryption of recorded traffic, meaning that each quantum-vulnerable handshake permanently increases future confidentiality risk. By contrast, signature forgery attacks require active exploitation with a CRQC at the time of connection and do not retroactively compromise previously captured sessions.

For this reason, PQ key establishment has emerged as the primary near-term focus of PQC transition efforts for TLS.

%% file: tables/2_shortlisted_countries.tex
% \usepackage{booktabs}
\begin{table}[t]
\renewcommand{\arraystretch}{1.5}

    \caption{Shortlisted countries and their regions}
    \label{tab:2_shortlisted_countries}
    
    \centering
    \small

    \resizebox{\columnwidth}{!}{
        \rowcolors{2}{}{lightgray}    
        
        \begin{tabular}{*{1}{L{1.8cm}}*{1}{L{7cm}}}

            \toprule
            
            \textbf{Region} & 
            \textbf{Countries} \\

            \midrule
            
            North America & 
            United States, Canada, Mexico \\

            \cmidrule(){1-2}

            \rowcolor{white}
            Europe & 
            France, Germany, United Kingdom, Netherlands, Switzerland, Belgium, Spain, Italy, Denmark, Norway, Czech Republic, Russia, Ukraine \\

            \cmidrule(){1-2}
            
            Asia & 
            China, Japan, South Korea, Singapore, Taiwan, Israel, India, Indonesia, Turkiye, Saudi Arabia \\

            \cmidrule(){1-2}

            \rowcolor{white}
            Oceania & 
            Australia, New Zealand \\

            \cmidrule(){1-2}

            South America & 
            Brazil, Argentina \\

            \cmidrule(){1-2}
            
            \rowcolor{white}            
            Africa & 
            Senegal \\        

            \bottomrule
    
        \end{tabular}
    }
\end{table}

%% file: tables/3_domcop_traco_crux_summary.tex
\begin{table}[t]
\renewcommand{\arraystretch}{1}

    \caption{TLS scanning outcomes for the three domain ranking datasets used in our robustness analysis.}
    \label{tab:3_domcop_traco_crux_summary}
    
    \centering
    \scriptsize
    
    \rowcolors{2}{}{lightgray}
        \begin{tabular}{lccc}
        
            \toprule
            
                Dataset & Domains Scanned & Successful & Fail \\
                
            \midrule
                DomCop & 1,000,000 & 714,067 & 285,933 \\
            \cmidrule(){1-4}
            \rowcolor{white}
                Tranco & 1,000,000 & 669,810 & 330,190 \\
            \cmidrule(){1-4}
                CrUX   &   999,888 & 773,332 & 226,556 \\
            \bottomrule
            
        \end{tabular}
    
\end{table}

%% file: tables/4_domcop_tranco_crux_adoption.tex
\begin{table}[t]
\renewcommand{\arraystretch}{1.5}
    \caption{PQ-TLS adoption across domain ranking datasets (March 2026)}
    \label{tab:4_domcop_tranco_crux_adoption}

    \centering
    \resizebox{\columnwidth}{!}{
        \rowcolors{2}{}{lightgray}
        \begin{tabular}{*{1}{L{3.2cm}}*{6}{M{1.2cm}}}

            \toprule

            \multicolumn{1}{c}{\multirow{2}{*}{\textbf{PQC KEM}}} &
            \multicolumn{3}{c}{\textbf{Default}} &
            \multicolumn{3}{c}{\textbf{Support}} \\
            \cmidrule(lr){2-4} \cmidrule(lr){5-7}
            \rowcolor{white}
            &
            \textbf{DomCop} &
            \textbf{Tranco} &
            \textbf{CrUX} &
            \textbf{DomCop} &
            \textbf{Tranco} &
            \textbf{CrUX} \\
            \midrule
            \multicolumn{7}{l}{\textit{Hybrid KEMs}} \\
            X25519MLKEM768     & 49.74\% & 51.55\% & 52.75\% & 49.75\% & 51.56\% & 52.75\% \\
            SecP256r1MLKEM768  & 0.00\%  & 0.00\%  & 0.00\%  & 5.44\%  & 4.07\%  & 6.88\% \\
            SecP384r1MLKEM1024 & 0.00\%  & 0.00\%  & 0.00\%  & 1.91\%  & 0.40\%  & 0.24\% \\

            \midrule

            \multicolumn{7}{l}{\textit{Pure PQC KEMs}} \\
            ML-KEM-512  & 0.00\% & 0.00\% & 0.00\% & 0.00\% & 0.00\% & 0.00\% \\
            ML-KEM-768  & 0.00\% & 0.00\% & 0.00\% & 0.00\% & 0.00\% & 0.00\% \\
            ML-KEM-1024 & 0.00\% & 0.00\% & 0.00\% & 0.64\% & 0.12\% & 0.33\% \\

            \bottomrule

        \end{tabular}
    }
\end{table}

%% file: figures/4_all_sectors/4_all_sectors_figure.tex
% Wrapper name differs from the PDF basename so arXiv preserves the figure asset.
\begin{figure*}[!t]
    \centering
    \includegraphics[width=\textwidth]{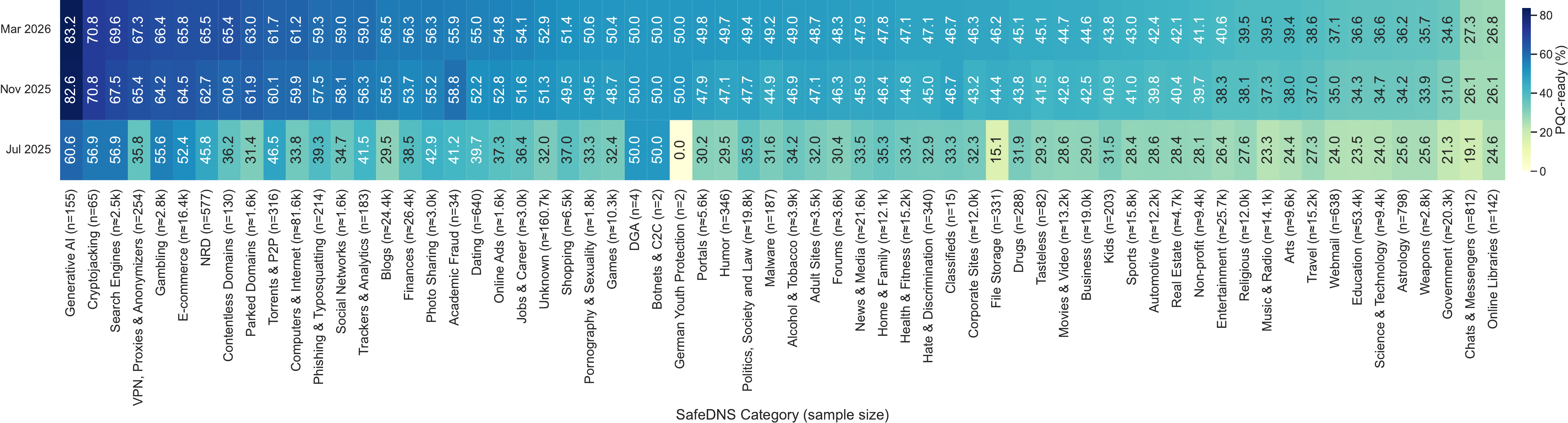}

    \caption{PQ-TLS Adoption Trends Across All SafeDNS Categories}

    \label{fig:4_all_sectors}    
    
\end{figure*}